\documentclass[a4paper,11pt]{article}
\pdfoutput=1 
\usepackage{jheppub} 
\usepackage{amsmath,amsfonts,amssymb,mathrsfs,graphicx,color,longtable,bm}
\usepackage{hyperref}
\usepackage{graphicx}
\usepackage{array}
\usepackage{color}
\usepackage{enumitem}
\usepackage[usenames,dvipsnames,table]{xcolor}
\usepackage[T1]{fontenc} 
\usepackage[utf8]{inputenc}
\usepackage{amsmath,amssymb,epsfig,txfonts,relsize,xspace}

\allowdisplaybreaks

\def\etmiss{\ensuremath{E_{T}^{\mathrm{miss}}}\xspace}
\def\ptmiss{\ensuremath{\vec p^{\mathrm{\ miss}}_T\xspace}}

\newcommand{\ttbar}{\ensuremath{t\bar{t}}\xspace}

\def\TeV{\ifmmode {\mathrm{\ Te\kern -0.1em V}}\else
                   \textrm{Te\kern -0.1em V}\fi}
\def\GeV{\ifmmode {\mathrm{\ Ge\kern -0.1em V}}\else
                   \textrm{Ge\kern -0.1em V}\fi}
 
\definecolor{nicered}{rgb}{0.7,0.1,0.1}
\definecolor{nicegreen}{rgb}{0.1,0.5,0.1}
\definecolor{niceblue}{rgb}{0.0,0.1,0.7}
\hypersetup{colorlinks,citecolor=nicegreen,linkcolor=nicered,urlcolor=niceblue}
                                      
\def\bm#1{\mbox{\boldmath$#1$\unboldmath}}
\def \beq{\begin{equation}}
\def \eeq{\end{equation}}
\def \bea{\begin{eqnarray}}
\def \eea{\end{eqnarray}}

\begin{document}

\title{Resonant third-generation leptoquark \\ signatures  at the Large Hadron Collider}

\author[1]{Ulrich Haisch}

\author[2]{and Giacomo Polesello}

\affiliation[1]{Max Planck Institute for Physics, F{\"o}hringer Ring 6,  80805 M{\"u}nchen, Germany}

\affiliation[2]{INFN, Sezione di Pavia, Via Bassi 6, 27100 Pavia, Italy}

\emailAdd{haisch@mpp.mpg.de}
\emailAdd{giacomo.polesello@cern.ch}

\abstract{Given the hints of lepton-flavour non-universality  in $B$-meson decays,  leptoquarks~(LQs)  are enjoying a renaissance. We propose novel Large Hadron Collider~(LHC) searches for such hypothetical states that do not rely on  strong production only, but  can also receive important contributions from quark-lepton annihilation. For the cases of a resonant signal involving a bottom quark and a tau lepton ($b + \tau$), a top quark and missing transverse energy~($\etmiss$) and light-flavour jets plus $\etmiss$, we develop realistic analysis strategies and provide detailed evaluations of the achievable sensitivities for the corresponding LQ signatures at future LHC runs. Our analyses allow us to derive a series of stringent constraints on the masses and couplings of third-generation singlet  vector~LQs, showing that at LHC~Run~III and the high-luminosity LHC the proposed search channels  can probe interesting parts of the LQ parameter space addressing the~$B$-physics anomalies. In view of the reach of the proposed $b + \tau$ signature, we recommend that dedicated resonance searches for this final state should be added to the exotics search canon of both ATLAS and CMS.}

\maketitle
\flushbottom

\section{Introduction}
\label{sec:introduction}

The deviations from  $\tau$-$\mu$ (and $\tau$-$e$)  universality in $b \to c \ell \nu$ transitions~\cite{Lees:2012xj,Huschle:2015rga,Aaij:2015yra,Sato:2016svk,Hirose:2016wfn,Aaij:2017uff} and the deviations from~$\mu$-$e$  universality in $b \to s \ell^+ \ell^-$ transitions~\cite{Aaij:2014ora,Aaij:2017vbb} are commonly considered the most compelling departures from the Standard Model (SM) observed  by collider  experiments in recent years. As a result of a significant amount of theoretical work~\cite{Alonso:2015sja,Calibbi:2015kma,Fajfer:2015ycq,Barbieri:2015yvd,Hiller:2016kry,Bhattacharya:2016mcc,Barbieri:2016las,Buttazzo:2017ixm,Assad:2017iib,DiLuzio:2017vat,Calibbi:2017qbu,Bordone:2017bld,Barbieri:2017tuq,Blanke:2018sro,Greljo:2018tuh,Bordone:2018nbg,Kumar:2018kmr,Azatov:2018kzb,DiLuzio:2018zxy,Angelescu:2018tyl,Schmaltz:2018nls,Fornal:2018dqn,Aebischer:2019mlg,Cornella:2019hct,Shi:2019gxi,DaRold:2019fiw,Bordone:2019uzc,Crivellin:2019dwb,Altmannshofer:2020ywf,Iguro:2020keo} it is by now well-established that singlet vector leptoquarks (LQs) with a mass in the TeV range and third-generation couplings provide a simple, especially appealing explanation of  both sets of anomalies. 

Several different search strategies for third-generation LQs have so far been considered at the Large Hadron Collider~(LHC). While the ATLAS and the CMS collaborations have mainly focused on LQ  pair production via strong interactions in gluon-gluon fusion or quark-antiquark annihilation (cf.~\cite{Sirunyan:2018ruf,Sirunyan:2018vhk,Aaboud:2019bye,Aad:2020jmj,cms_2012.04178} for the latest results), the importance of LQ pair production via $t$-channel exchange of a lepton, of LQ exchange  in Drell-Yan~(DY) like di-lepton production and of single LQ production in gluon-quark fusion in constraining the quark-lepton-LQ couplings has  also been established~\cite{Mandal:2015vfa,Dorsner:2016wpm,Faroughy:2016osc,Raj:2016aky,Dorsner:2018ynv,Hiller:2018wbv,Bansal:2018eha,Sirunyan:2018jdk,Angelescu:2018tyl,Schmaltz:2018nls,Mandal:2018kau,Baker:2019sli,Chandak:2019iwj,Borschensky:2020hot,Brooijmans:2020yij,Bhaskar:2020gkk}. 

Due to quantum fluctuations, protons however also contain charged leptons, making it is possible to target lepton-induced processes at the LHC as well. The simplest process of this kind consists in the collision between a quark from one proton  and a lepton from the other proton, giving rise to resonant single LQ production at hadron colliders~\cite{Ohnemus:1994xf}.  In fact, using the precise determination of the lepton parton distribution functions~(PDFs) obtained recently in~\cite{Buonocore:2020nai}, it has been shown in~\cite{Buonocore:2020erb} that LHC searches for $s$-channel single LQ production provide the strongest constraints to date  on all the flavour combinations of first- and second-generation minimal scalar LQs that involve an up or a down quark. Given the suppression of the relevant quark PDFs, the constraints on minimal scalar LQ interactions involving a strange or a charm quark turn out to be less stringent but still relevant in view of the large amount of data collected in the high-luminosity era of the LHC~(HL-LHC). 

The main goal of this article is to extend and generalise the basic ideas and results presented in the publication~\cite{Buonocore:2020erb} (see subsequently also~\cite{Greljo:2020tgv}) to the case of singlet vector LQs, coupled  mainly to third-generation fermions. To this purpose, we develop three  search strategies for LQ signatures that are induced by quark-lepton annihilation at the LHC. The~first search strategy exploits final states with a bottom quark~($b$) and a tau lepton~($\tau$), while the second and third search strategy targets final states with a single top quark~($t$) and  significant amounts of missing transverse energy~($\etmiss$) and light-flavour~jets~($j$) plus~$\etmiss$, respectively. In all three cases we provide detailed evaluations of the achievable sensitivities for the  LQ signature at LHC~Run~III and the HL-LHC. These sensitivities  are then used to constrain  singlet vector LQ models with  left- as well as third-generation right-handed quark-lepton-LQ couplings and/or sizeable second-third-generation left-handed mixing.  LQ scenarios of this kind have also been considered in the recent detailed collider analysis~\cite{Baker:2019sli}. We~will benchmark the  bounds derived in our work against the limits obtained in the latter~study. 

The remainder of this article is organised as follows. In Section~\ref{sec:preliminary} we briefly describe the structure of the LQ model that we consider, while we explain  in Section~\ref{sec:motivation} which  resonant LQ production channels are particularly motivated by the $B$-physics anomalies. The main ingredients of our Monte~Carlo~(MC) generation and our detector simulation are discussed in Section~\ref{sec:MC}.  The actual analysis strategies are detailed in~Section~\ref{sec:strategies}. In~Section~\ref{sec:results} we present our numerical results and examine the  sensitivity of the studied LQ signatures at upcoming LHC runs. We~conclude and present an outlook in~Section~\ref{sec:conclusions}.

\section{Theoretical framework}
\label{sec:preliminary}

In  this article, we consider a LQ called $U$ that transforms as $(\bm{3},\bm{1})_{2/3}$  under the SM gauge group $S\!U(3)_c \times S\!U(2)_L \times U(1)_Y$. Such a LQ can either appear as a massive gauge boson of a  spontaneously broken gauge symmetry~\cite{Assad:2017iib,DiLuzio:2017vat,Calibbi:2017qbu,Bordone:2017bld,Barbieri:2017tuq,Blanke:2018sro,Greljo:2018tuh,Bordone:2018nbg,DiLuzio:2018zxy,Fornal:2018dqn,Cornella:2019hct} or arise as  a massive vector resonance of some new strongly-interacting dynamics~\cite{Barbieri:2015yvd,Barbieri:2016las,Buttazzo:2017ixm,DaRold:2019fiw}. Irrespectively of its ultraviolet~(UV) origin, the quantum numbers of $U$  unambiguously fix the quark-lepton-LQ interactions to have the form 
\beq \label{eq:LU}
{\cal L} \supset \frac{g_U}{\sqrt{2}} \left [  \beta^{ij}_L \, \bar Q^{i, a}_L \gamma_\mu L^j_L + \beta^{ij}_R \, \bar d^{i, a}_R \gamma_\mu \ell^j_R  \right ]  U^{\mu, a} \ +{\rm h.c.}  \,, 
\eeq
where, without loss of generality, the down-type quark and charged-lepton mass eigenstate basis has been adopted for the  left-handed fermion multiplets:
\beq \label{eq:fermiondoublets}
Q_L^i = \begin{pmatrix} V_{ji}^\ast \hspace{0.5mm} u_L^j \\ d_L^i \end{pmatrix} \,, \qquad 
L_L^i =  \begin{pmatrix}  \nu_L^i \\ \ell^i_L \end{pmatrix} \,.
\eeq
In~(\ref{eq:LU}) and~(\ref{eq:fermiondoublets}) the elements of the Cabibbo-Kobayashi-Maskawa~(CKM) matrix are denoted by $V_{ij}$, the fields $d_R$~($\ell_R$) represent the right-handed down-type quark~(charged-lepton) singlets, $i,j \in \{1,2,3\}$ are flavour indices, $a \in \{ 1, 2, 3 \}$ is a  colour index,~$g_U$ denotes the overall coupling strength of the quark-lepton-LQ interactions, and $\beta^{ij}_{L,R}$ are complex~$3 \times 3$ matrices in flavour space. 

The latest global analyses~(see for instance~\cite{Aebischer:2019mlg,Baker:2019sli,Cornella:2019hct}) of   lepton-flavour universality~(LFU) violation in charged-current $b \to c$ transitions $\big($i.e.~$R_{D^{(\ast)}}$$\big)$ and in neutral-current $b \to s$ transitions $\big($i.e.~$R_{K^{(\ast)}}$$\big)$ show that a singlet vector LQ with a mass $M_U = {\cal O} (1 \, {\rm TeV})$ and a coupling $g_U = {\cal O} (1)$ provides an excellent description of the $B$ anomalies if the left-handed quark-lepton-LQ couplings entering~(\ref{eq:LU}) have the following hierarchy: 
\beq \label{eq:betaL}
\left | \beta^{22}_L \right | \lesssim \left |  \beta^{32}_L  \right | \ll \left | \beta^{23}_L \right | \lesssim \left |  \beta^{33}_L \right | = {\cal O} (1) \,.
\eeq
While there is no clear indication of right-handed currents in present $B$-meson data, motivated UV completions of~(\ref{eq:LU}) such as those proposed in~\cite{Bordone:2017bld,Bordone:2018nbg,Cornella:2019hct} give rise to: 
\beq \label{eq:betaR}
 \left |  \beta^{33}_R \right | = {\cal O} (1) \,.
\eeq
In order to restrict the number of model parameters we furthermore assume that all quark-electron-LQ couplings vanish, and that $\beta^{13}_L = V_{td}^\ast/V_{ts}^\ast  \, \beta^{23}_L \simeq -\lambda \, \beta^{23}_L$ as in models~\cite{Barbieri:2015yvd,Buttazzo:2017ixm,Bordone:2017bld,Barbieri:2017tuq,Greljo:2018tuh,Bordone:2018nbg,DiLuzio:2018zxy,Baker:2019sli,Cornella:2019hct} based on a minimal breaking of the assumed $U(2)_Q$ flavour symmetry of first- and second-generation quarks. Here $\lambda \simeq 0.2$ denotes the Cabibbo angle. Both assumptions are phenomenologically well-motivated given the tight  constraints from low-energy measurements, in particular the bounds  on lepton-flavour violation in charged-lepton decays and the limits on neutral meson mixing (see~e.g.~\cite{Buttazzo:2017ixm,Bordone:2018nbg,Cornella:2019hct}). 

\section{Resonant LQ  signals motivated by $\bm{B}$  anomalies}
\label{sec:motivation}

The discussion in the previous section should have made clear that  in the singlet vector LQ model a successful explanation of the $B$-physics anomalies relies largely on the five variables
\beq \label{eq:input}
M_U \,, \qquad 
g_U \,, \qquad 
\beta^{33}_L \,, \qquad 
\beta^{23}_L \,,  \qquad 
\beta^{33}_R \,,
\eeq
while the left-handed couplings $\beta^{32}_L$ and $\beta^{22}_L$ represent only  small perturbations needed to accommodate the deviations observed in the $b \to s$ sector.  Hereafter we will therefore focus on the subset~(\ref{eq:input}) of model parameters, setting the left-handed couplings  $\beta^{32}_L$ and $\beta^{22}_L$ as well  as all right-handed couplings other than $\beta^{33}_R$ to zero. 

\begin{table}
\def\arraystretch{1.25}
\begin{center}
\begin{tabular}{|ccc|cccc|}
\hline 
\multicolumn{3}{|c|}{Parameters} & \multicolumn{4}{c|}{Branching ratios} \\ 
$\beta^{33}_L$ & $\beta^{23}_L$ & $\beta^{33}_R$ & ${\rm BR} \left (U \to b \tau^+ \right  )$ & ${\rm BR} \left (U \to t \bar \nu_\tau \right  )$  & ${\rm BR} \left (U \to s \tau^+ \right  )$ & ${\rm BR} \left (U \to c \bar \nu_\tau \right  ) $ \\[1mm]
\hline 
1 & 0 & 0 & 51\% & 49\% & 0\% & 0\% \\
1 & 1 & 0 & 25\% & 22\% & 25\% & 27\% \\
1 & 0 & 1 & 68\% & 32\% & 0\% & 0\% \\
\hline
 \end{tabular}
\vspace{2mm}
\caption{Branching ratios of $U$ for $M_U = 1 \, {\rm TeV}$ and three different choices of $\beta^{33}_L$, $\beta^{23}_L$ and $\beta^{33}_R$. }
\label{tab:BRs}
\end{center}
\end{table}

In the limit  of $M_U \gg m_f$ with $m_f$ denoting the masses of the SM fermions and working to leading order in the Cabibbo angle, the tree-level expressions of the   relevant partial decay widths of  the singlet vector LQ read 
\beq \label{eq:GammaU} 
\begin{split}
\Gamma \left (U \to b \tau^+ \right  ) & \simeq \frac{g_U^2}{48 \pi}  \left ( \left |\beta^{33}_L \right |^2 + \left |\beta^{33}_R \right |^2 \right )  M_{U} \,, \qquad 
\Gamma \left (U \to t \bar \nu_\tau \right  )  \simeq \frac{g_U^2}{48 \pi}  \left |\beta^{33}_L \right |^2  M_{U} \,, \\[2mm]
\Gamma \left (U \to s \tau^+ \right  ) & \simeq \frac{g_U^2}{48 \pi}  \left |\beta^{23}_L \right |^2  M_{U} \,, \hspace{2.25cm}
\Gamma \left (U \to c \bar \nu_\tau \right  )  \simeq  \frac{g_U^2}{48 \pi}  \left |\beta^{23}_L \right |^2  M_{U} \,.
\end{split}
\eeq
In Table~\ref{tab:BRs} we report the corresponding branching ratios of $U$ for $M_U = 1\, {\rm TeV}$ and three different choices of the parameters $\beta^{33}_L$, $\beta^{23}_L$ and $\beta^{33}_R$ --- notice that due to phase-space  and quark-mixing effects the shown branching ratios deviate slightly from the values expected from~(\ref{eq:GammaU}). From the above formulas one observes that  for fixed $\beta^{33}_L$ increasing $\beta^{23}_L$ increases the $j + \etmiss$~(i.e.~mono-jet) signal strength relative to $b + \tau$ and $t + \etmiss$~(i.e.~mono-top) and vice versa.  Varying the coupling~$\beta^{33}_R$ instead allows to change the  importance of the $b + \tau$ final state relative to the $t + \etmiss$ and  $j + \etmiss$  final states with larger (smaller) values of $\beta^{33}_R$ enhancing (decreasing) the $b + \tau$ branching ratio. 

\begin{figure}[!t]
\begin{center}
\vspace{2mm} 
\includegraphics[width=0.8\textwidth]{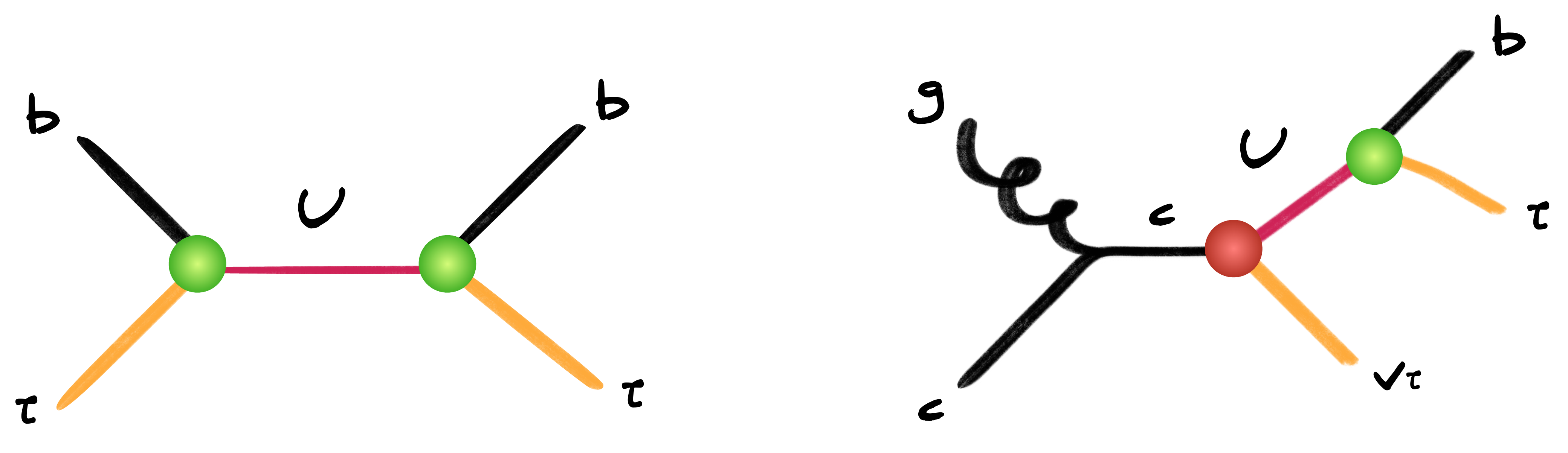} 
\vspace{4mm} 
\caption{\label{fig:diagrams} Examples of tree-level Feynman diagrams giving rise to a final state involving a $b$ quark and a $\tau$ lepton. The green vertices indicate the couplings $\beta^{33}_L$ or  $\beta^{33}_R$, while the red vertex corresponds to  $\beta^{23}_L$. Graphs with $t$-channel exchange of a $U$ also contribute to $b \tau \to b \tau$ scattering if  $\beta^{33}_L \neq 0$ or $\beta^{33}_R \neq 0$,  while the $s$-channel process $s \tau \to b \tau$ is also possible if $\beta^{23}_L \neq 0$.  Diagrams corresponding to the latter two $2 \to 2$ transitions are not depicted. In the case of the $2 \to 3$ process also diagrams with $t$-channel exchange of a~$U$ and graphs involving a gluon-LQ-LQ vertex contribute to  the shown $gc \to b \tau \nu_\tau$ reaction. These contributions have again not been displayed. For further details consult the main text.  }
\end{center}
\end{figure}

The relative magnitudes of the couplings $\beta^{33}_L$, $\beta^{23}_L$ and $\beta^{33}_R$ do not only dictate the decay pattern of the singlet vector LQ, but also determine how the $U$ is produced. This feature is illustrated in~Figure~\ref{fig:diagrams}, which shows examples of tree-level diagrams leading to resonant single $U$ production via bottom-tau annihilation~(left) and single $U$ production in gluon-charm annihilation~(right) followed by the decay $U \to b \tau$. Hereafter we will for simplicity often refer to the two different production mechanisms as the $2 \to 2$ and the $2 \to 3$ process, respectively.  As indicated in the figure by the coloured vertices for $\beta^{33}_L \neq 0$ or $\beta^{33}_R \neq 0$ the process $b \tau \to U \to b \tau$ takes place even if  $\beta^{23}_L = 0$, while the rate for $g c \to  U \nu_\tau \to b \tau \nu_\tau$ is non-zero only if  $\beta^{23}_L \neq 0$. In fact, in view of~(\ref{eq:betaL}) and~(\ref{eq:betaR}) one expects  that singlet vector LQ model realisations that provide an explanation of the $R_{D^{(\ast)}}$ anomalies feature resonant $b + \tau$ and mono-top production in bottom-tau annihilation at the LHC. Mono-jet production via $b \tau \to c \nu_\tau$ or $s \tau  \to c \nu_\tau$  is also possible but requires the second-third-generation left-handed mixing parameter $\beta^{23}_L$ to be sufficiently large. 

The above discussion of the decay pattern and the production mechanisms of $U$  singles out the~$b + \tau$, $t + \etmiss$ and $j + \etmiss$ final states as the most promising resonant production channels to search for the presence of a singlet vector  LQ as motivated by the $b \to c \, (s)$ anomalies. Other important search channels  for third-generation singlet vector LQs are pair production leading to $b \bar b \tau^+ \tau^-$ and $t \bar t \nu_\tau \bar \nu_\tau$ final states, DY-like $\tau^+ \tau^-$~(i.e.~di-tau) production as well as  $\tau \nu_\mu$ and~$\tau \mu$ production through $t$-channel exchange of a $U$ --- for a comprehensive analysis  of these final states see~\cite{Baker:2019sli}. In~the following, we will concentrate on the $b + \tau$, mono-top and mono-jet signatures because the LHC sensitivity of these  channels  to  the five-dimensional parameter space~(\ref{eq:input}) has not yet been~studied. 

\section{MC generation and detector simulation}
\label{sec:MC}

The signal predictions are calculated at leading order~(LO) using the implementation~\cite{Baker:2019sli} of the Lagrangian~(\ref{eq:LU}) together with  the {\tt LUXlep}~PDFs, which have been obtained by combining the  lepton PDFs of~\cite{Buonocore:2020nai} with the  {\tt NNPDF3.1luxQED} set~\cite{Bertone:2017bme}. The generation and showering of the LQ samples is  performed  with {\tt MadGraph5\_aMC\@NLO}~\cite{Alwall:2014hca} and {\tt PYTHIA~8.2}~\cite{Sjostrand:2014zea}, respectively. Since {\tt PYTHIA~8.2} presently cannot deal with incoming leptons, all initial state leptons (i.e.~$e$,~$\mu$ and~$\tau$)  have been replaced by photons in the Les~Houches files before showering the events. Our signal simulations therefore do not include leptons but quarks arising from photon splitting in the parton shower~(PS) backward evolution. We expect the resulting  mismodelling of the hadronic and leptonic activity of the LQ signals to have only a  very minor impact on the numerical results obtained below in~Section~\ref{sec:results}.

In the case of the $b + \tau$ and the mono-top signature all SM processes that contain  one or two charged leptons from the decay of a electroweak~(EW) gauge boson $V = W,Z$ or the decay of a~$\tau$ lepton are included in the background. The generation of the relevant $b + \tau$ and mono-top backgrounds follows~\cite{Haisch:2018bby}. Specifically, the backgrounds from $\ttbar$~\cite{Campbell:2014kua}, $tW$~\cite{Re:2010bp}, $WW$, $WZ$ and $ZZ$ production~\cite{Melia:2011tj,Nason:2013ydw} are generated at next-to-leading order~(NLO) in~QCD with {\tt POWHEG~BOX}~\cite{Alioli:2010xd}. The $V + {\rm jets}$ backgrounds are generated at LO using {\tt  MadGraph5\_aMC@NLO} and include up to four additional jets. The $\ttbar V$ backgrounds  are also simulated  at LO with {\tt  MadGraph5\_aMC@NLO} and include up to two additional jets, while the $tZ$ and $tWZ$ backgrounds are obtained at LO with the same MC generator. The production of $b + \tau$  from an initial-state bottom quark and a tau lepton via $t$-channel exchange of a photon or $Z$ boson also represents an irreducible background. We include this background at~LO using {\tt  MadGraph5\_aMC@NLO}. All partonic events are showered with {\tt PYTHIA~8.2}. The samples produced with {\tt POWHEG~BOX} are normalised to the corresponding NLO QCD cross sections, except for $t\bar{t}$ which is normalised to the  cross section obtained at next-to-next-to-leading order (NNLO) in~QCD plus next-to-next-to-leading logarithmic QCD corrections~\cite{Czakon:2011xx,Czakon:2013goa}. The $V + {\rm jets}$ samples are normalised to the NNLO QCD cross sections~\cite{Anastasiou:2003ds,Gavin:2012sy} and the~$\ttbar V$~samples are normalised to the NLO QCD cross section as calculated by  {\tt  MadGraph5\_aMC@NLO}.

For the mono-jet signature, the dominant SM backgrounds arise from $V +{\rm jets}$ production. The only relevant process not included in the one-lepton backgrounds described above is the $Z+\mathrm{jets}$ channel followed by the decay $Z \to \nu \bar \nu$. Like in  our earlier work~\cite{Haisch:2018hbm} it is   generated at LO with {\tt  MadGraph5\_aMC@NLO}, and can contain up to two additional jets. The generation is performed in slices of the vector-boson~$p_T$, and the resulting events are showered with {\tt PYTHIA~8.2} employing   a Catani-Krauss-Kuhn-Webber jet-matching procedure~\cite{Catani:2001cc}. The inclusive signal region IM3  of the analysis~\cite{ATLAS-CONF-2020-048} requires $\etmiss > 350 \, {\rm GeV}$, and  for these selections the background from $V +{\rm jets}$ production amounts to around 95\% of the total SM background.  Our $V +{\rm jets}$  samples are normalised such that the different contributions match the number of events in the IM3 signal region as estimated by the ATLAS collaboration scaled to a  centre-of-mass~(CM) energy of $14 \, {\rm TeV}$ and to the appropriate integrated luminosity. The additional minor backgrounds from~$t\bar{t}$, $tW$ and diboson production are the same as in the $b + \tau$ and the mono-top case.

Electrons and  muons produced in the decays of real  EW gauge bosons and taus that are  isolated from  jets are considered in our analyses. Jets are built out of the moments of all the stable particles  depositing energy in the calorimeter except for muons using the anti-$k_t$ algorithm~\cite{Cacciari:2008gp} with a radius parameter of $R=0.4$, as implemented in {\tt FastJet}~\cite{Cacciari:2011ma}.  Jets originating from the hadronisation of bottom quarks ($b$-jets) and the hadronic decays of $\tau$ leptons are experimentally identified~(i.e.~tagged) with high efficiency. The \ptmiss \ vector with magnitude $\etmiss$  is constructed from  the transverse momenta of all the  invisible particles in the event. The experimental effects are simulated by smearing the momenta of the analysis objects and by applying efficiency factors where applicable. The used smearing and efficiency functions  are tuned to reproduce the performance of the  ATLAS detector~\cite{Aad:2008zzm,Aad:2009wy}.  In particular,  the performance of the ATLAS $b$-tagging algorithm is taken from~\cite{Aad:2019aic}. For the analyses  performed in this article,  a $b$-tagging working point is chosen that yields a  $b$-tagging efficiency of 77\%,  a  $c$-jet rejection  of 5 and a light-flavour jet rejection  of~110. The parametrisation of the $\tau$-tagging performance is taken from~\cite{ATL-PHYS-PUB-2019-005}. The used $\tau$-tagging working point  has an average efficiency of approximately 80\% and 70\%  for the identification of tau leptons in hadronic decays  into a single charge particle~(one-prong decays) and into three charged particles~(three-prong~decays), respectively. The assumed rejection factor for light-flavour  jets is taken to be 80 (500) for hadronic one-prong (three-prong) tau decays. 

\section{Analysis strategies}
\label{sec:strategies}

In this section we detail our analysis strategies that are designed to target the $b + \tau$, mono-top and mono-jet final states. For~each analysis strategy we spell out all selection criteria and illustrate their impact on the  SM background and the LQ  distributions of interest. 

\subsection[$b + \tau$ final state]{$\bm{b + \tau}$ final state}
\label{sec:btauanalysis}

The basic selection for the $b + \tau$ signature consists of  a $b$-tagged hadronic jet ($b$) and a hadronic jet corresponding to  the hadronic decay of a tau lepton ($\tau_{\rm had}$). We require  $p_T(b)>50 \, {\rm GeV}$ and $p_T(\tau_{\rm had})>150 \, {\rm GeV}$, where the latter requirement  is dictated by the expected trigger thresholds  for the single $\tau$ trigger at the HL-LHC~\cite{ATL-PHYS-PUB-2019-005}. Both the $b$-jet and the $\tau_{\rm had}$ are required to be within $|\eta|<2.5$, which is the pseudorapidity coverage of the ATLAS tracker. We further veto events with additional jets tagged as a $b$ or a $\tau$, or any additional light-flavour jet with $p_T>50 \, {\rm GeV}$. Moreover all  events containing a reconstructed light lepton ($e$ or $\mu$) are discarded.  Besides a $b$-jet and a~$\tau_{\rm had}$, the~$b + \tau$ signal also comprises $\etmiss$ associated to neutrinos. Notice that these neutrinos can either result from the~$\tau$ decay  itself  or from associated  production  depending on whether the~$2 \to 2$ or~the~$2 \to 3$ process is considered (see Figure~\ref{fig:diagrams}). Since the $2 \to 2$  and the $2 \to 3$ processes lead to  final states with very different kinematic features, it is essential to develop two separate strategies for them. While for LQ realisations~(\ref{eq:input}) with $\beta_L^{23} = 0$ only the strategy targeting the~$2 \to 2$ process is relevant, in all other cases both analyses can be applied. Furthermore, in the event that a LQ signal is observed, the relative rate in the two signal regions can be used to determine the composition of the signal which itself is controlled by the couplings $\beta_L^{33}$, $\beta_L^{23}$ and $\beta_R^{33}$. 

\begin{figure}[t!]
\begin{center}
\includegraphics[width=0.55\textwidth]{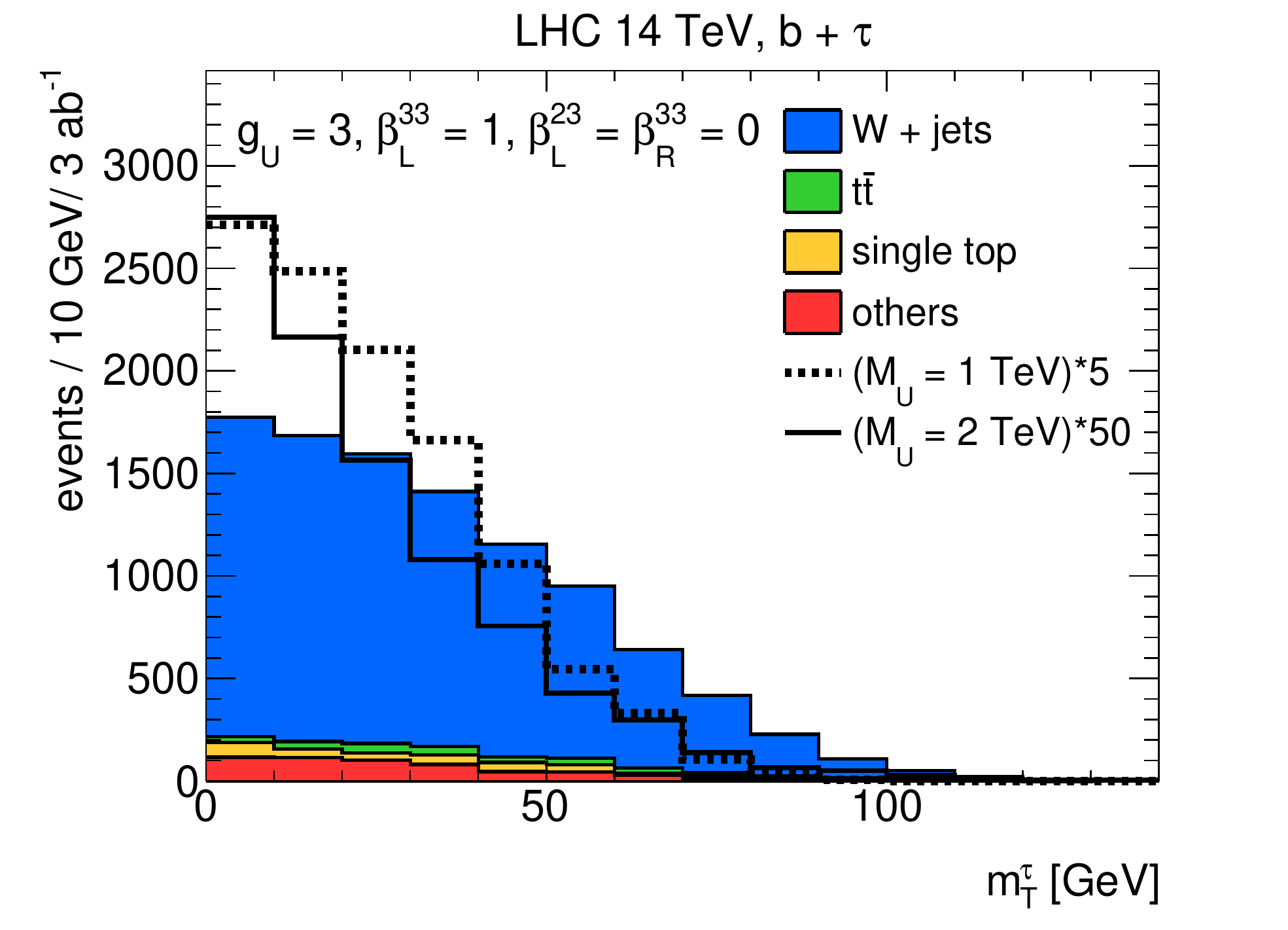}
\includegraphics[width=0.55\textwidth]{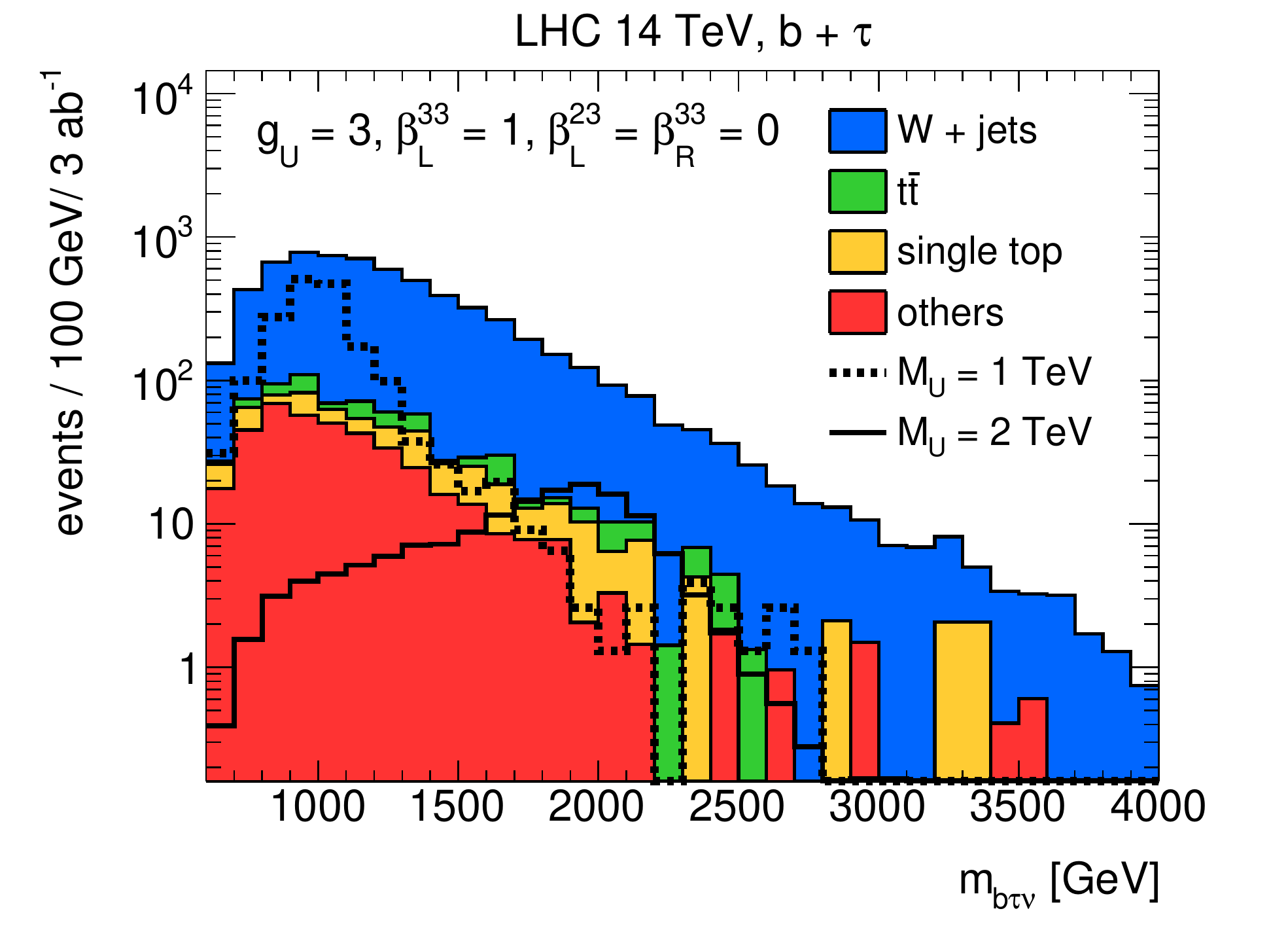}
\vspace{0mm}
\caption{Upper panel: $m_T^{\tau}$ distribution for the background and two $b + \tau$ signals scaled up by a factor of 5 and~50, respectively. Lower panel: $m_{b\tau\nu}$ distribution for the background  and two $b + \tau$  signals. The~background distributions (coloured histograms)  are stacked and the signal distributions in both panels correspond to  the LQ parameter choices $g_U = 3$, $\beta_L^{33} = 1$, $\beta_L^{23} = \beta_R^{33} = 0$ and  $M_U = 1 \, {\rm TeV}$ (black dotted lines) or $M_U = 2 \, {\rm TeV}$~(black solid lines). The shown predictions are obtained assuming LHC collisions at a~CM energy of~$14 \, {\rm TeV}$.}	
\label{fig:btau10}
\end{center}
\end{figure}

In the case of the  $2 \to 2$ process the entire amount of $\etmiss$ stems from the neutrino produced in the decay of the $\tau$. It follows that the transverse mass $m_T^{\tau}$ built from $\tau_{\rm had}$ and $\etmiss$   has an edge at the tau mass, which in practice is smeared by the experimental resolution on $\tau_{\rm had}$ and $\etmiss$. In~fact, the experimental distribution of $m_T^{\tau}$ displays a significant high-energy tail as well, corresponding to events where a semileptonic $B$-meson decay  occurs  inside the $b$-jet. In order to allow for a clean kinematic reconstruction of the $b$-$\tau_{\rm had}$-$\nu_\tau$ system we require  the azimuthal angle $\Delta\phi_{\tau \nu}$ between $\tau_{\rm had}$ and $\ptmiss$ to be smaller than 0.4, and $m_T^{\tau}<40 \, {\rm GeV}$, which efficiently select events  that contain only the tau neutrino from the hadronic decay of the $\tau$.  Notice that the cut on~$m_T^{\tau}$ in addition rejects SM events where a $\tau$ and a neutrino arise from a $W$-boson decay. This feature is illustrated in the upper panel of Figure~\ref{fig:btau10} which shows the $m_T^{\tau}$ distribution for the background and  two LQ parameter choices.   The displayed distributions are obtained after application  of the selection criteria given above plus the requirement $p_T(\tau_{\rm had})>300 \, {\rm GeV}$. An additional variable useful for the discrimination of signal and background is the ratio $\etmiss/p_T(\tau_{\rm had})$. For the $2 \to 2$  process all of the $\etmiss$ comes from the neutrinos from the $\tau$ decay, and therefore  $p_T(\tau_{\rm had})$ will be shared by the decay products depending on the mass and the orientation of the final-state particles. The~backgrounds are dominated by events where the $\tau$ is accompanied by an additional neutrino, which implies that  the ratio~$\etmiss/p_T(\tau_{\rm had})$ is typically higher for the signal than for the background.

\begin{figure}[t!]
\begin{center}
\includegraphics[width=0.55\textwidth]{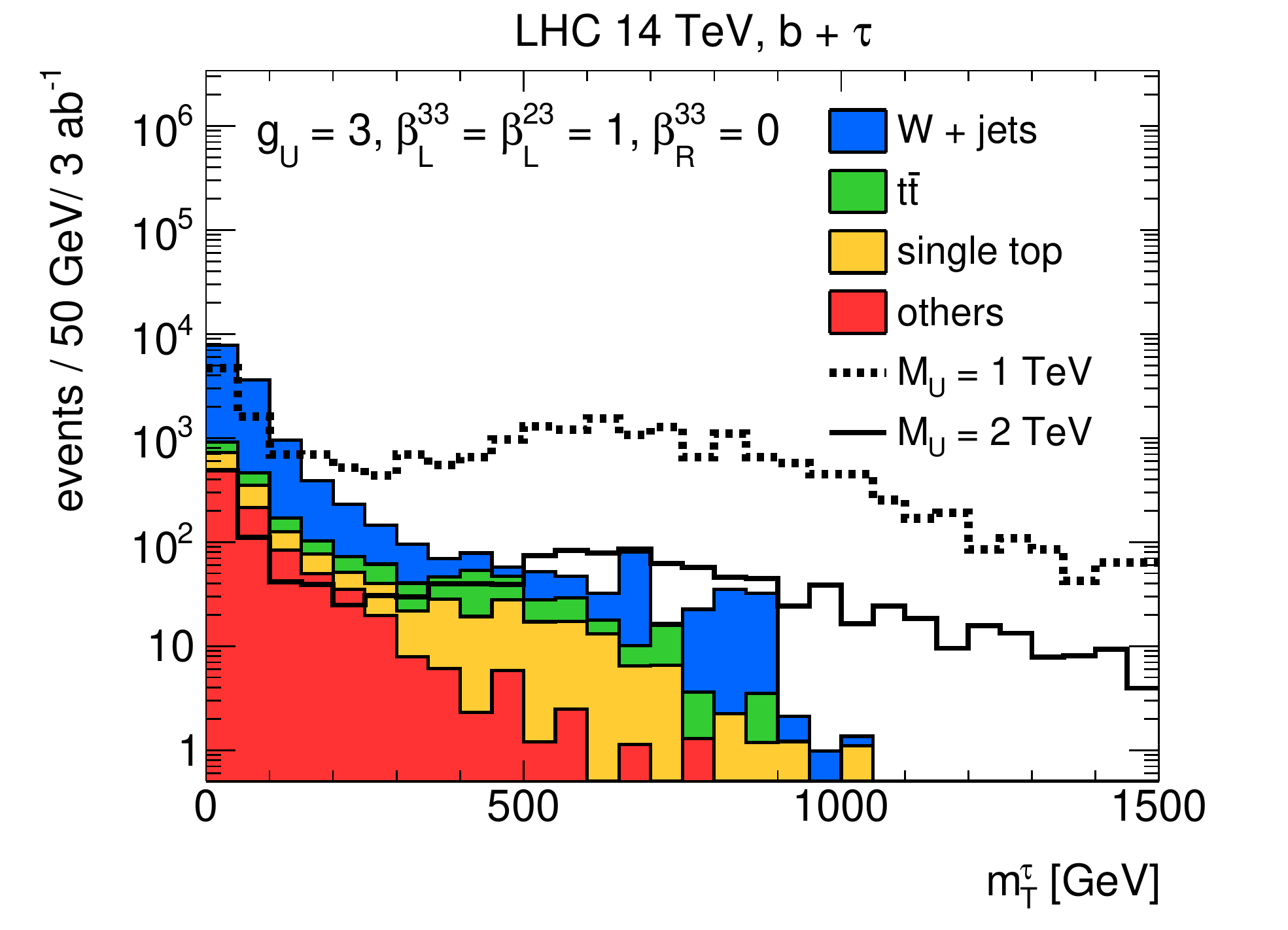}
\includegraphics[width=0.55\textwidth]{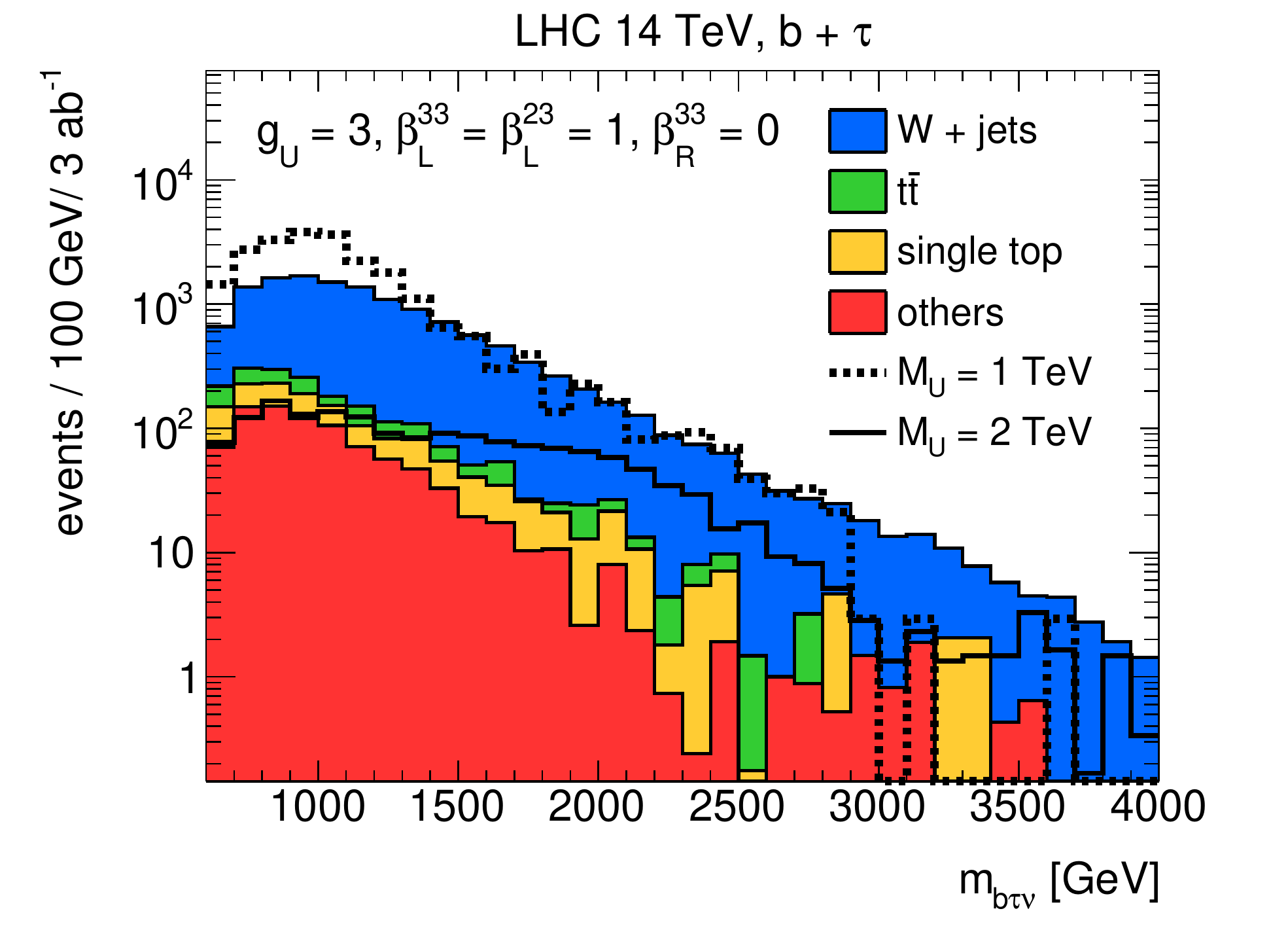}
\vspace{0mm}
\caption{As Figure~\ref{fig:btau10} but for  the LQ coupling choices $g_U = 3$, $\beta_L^{33} = \beta_L^{23} = 1$, $\beta_R^{33} = 0$.}
\label{fig:btau11}
\end{center}
\end{figure}

\begin{figure}[t!]
\begin{center}
\includegraphics[width=0.55\textwidth]{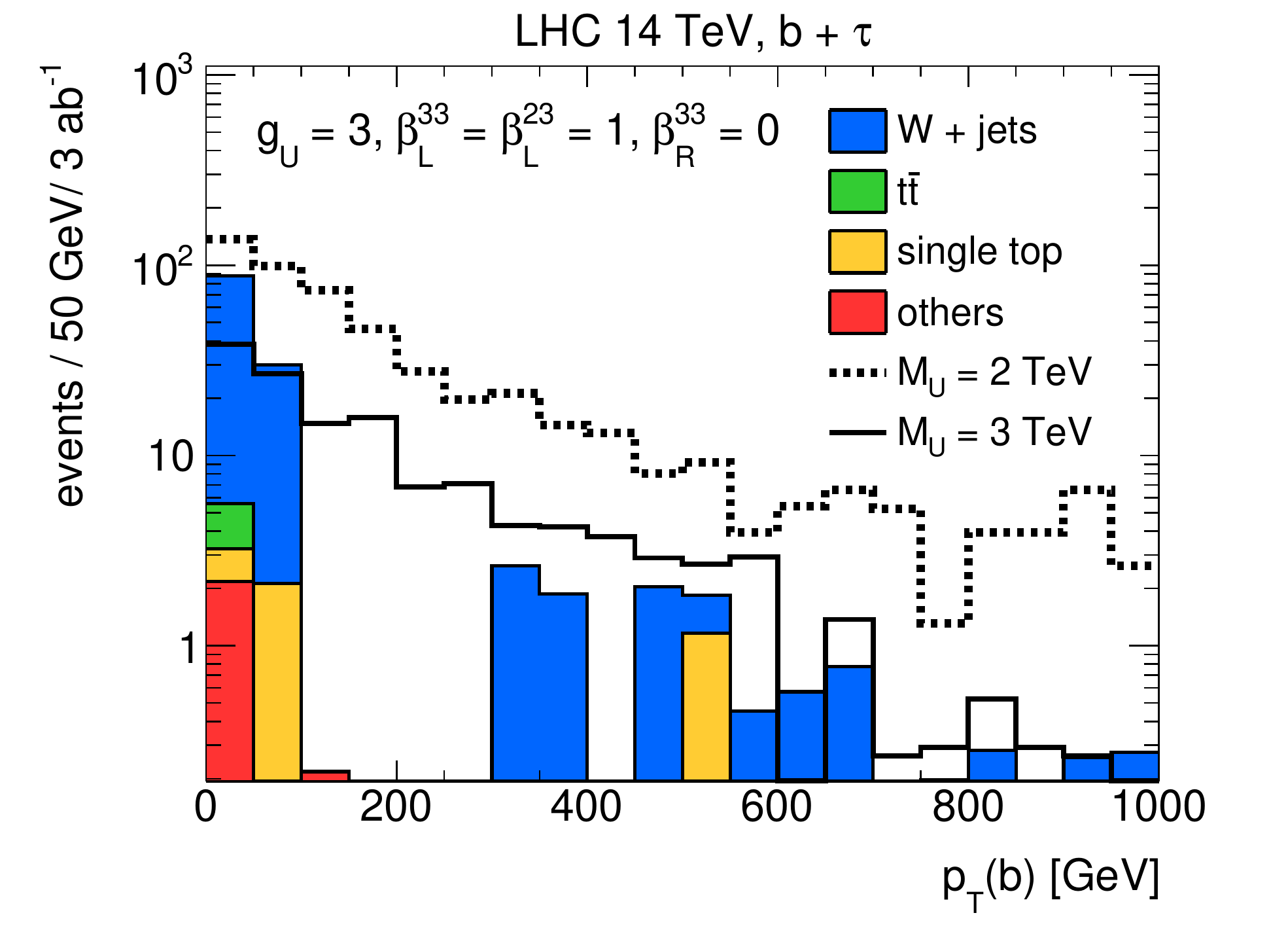} 
\includegraphics[width=0.55\textwidth]{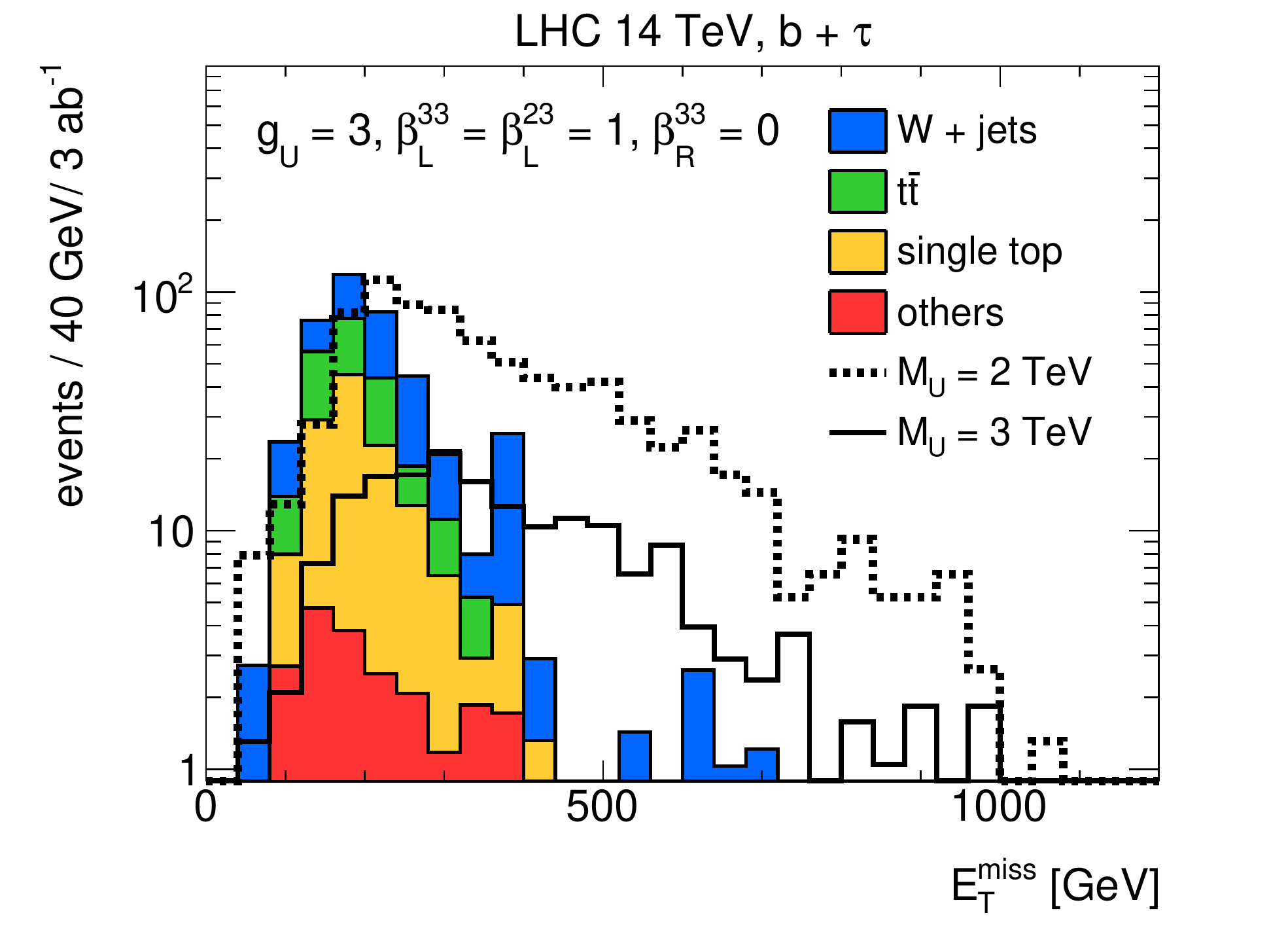} 
\includegraphics[width=0.55\textwidth]{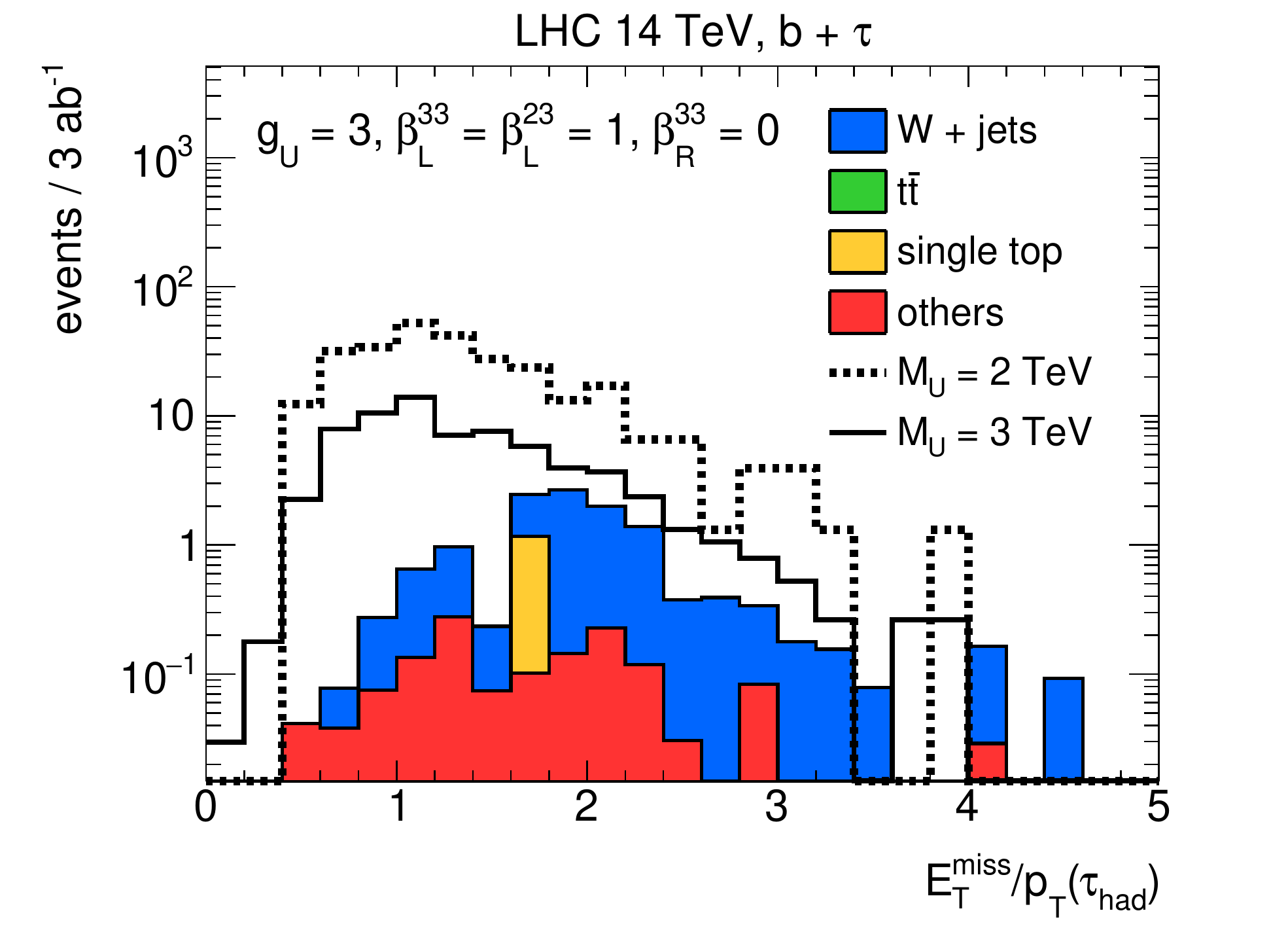} 
\vspace{0mm}
\caption{$p_T(b)$ (upper panel), $\etmiss$ (middle panel) and $\etmiss/p_T (\tau_{\rm had})$ (lower panel) distribution for the background and two $b + \tau$ signals after the selections for the $2 \to 3$ process as described in the main text. The displayed signal distributions correspond to $M_{U}=2 \, {\rm TeV}$~(black dotted lines)  and $M_U = 3 \, {\rm TeV}$~(black solid lines), while the other LQ~parameters  are the same as those used to obtain Figure~\ref{fig:btau11}.}
\label{fig:btau11add}
\end{center}
\end{figure}

For signal events where an on-shell LQ is produced, the~$\tau$~from the LQ decay will be highly boosted, and as a result  the momentum of the neutrino will tend to be aligned with the momentum of  the $\tau_{{\rm had}}$. In~this situation the invariant mass~$m_{b\tau\nu}$  of the $b$-jet and the $\tau$ can be fully reconstructed, and $m_{b\tau\nu}$  is predicted to peak at the LQ mass. In the lower panel of Figure~\ref{fig:btau10} we show the $m_{b\tau\nu}$ distribution for two LQ masses and the parameter choice  $\beta_L^{33} = 1, \beta_L^{23} = \beta_R^{33} = 0$, after imposing the cut $p_T(\tau_{\rm had})>300 \, {\rm GeV}$ in addition to the selections described above. One observes that the requirement $\left |m_{b\tau\nu} - M_U \right | <  200 \, {\rm GeV}$ with~$M_U$ denoting the nominal LQ mass efficiently selects the resonant decay of the LQ.  One also sees from this figure that the background is a  steeply falling distribution which is dominated by $W + {\rm jets}$ production. The final selection in the case of the  $2 \to 2$ process is obtained by imposing the latter requirement, by requiring $\etmiss/p_T(\tau_{\rm had})<0.3$ and by hardening the $p_T(\tau_{\rm had})$ threshold to a value between $300 \, {\rm GeV}$ and $600 \, {\rm GeV}$ depending on the LQ mass under consideration.

For the $2\rightarrow3$ process the additional  hard neutrino produced alongside the LQ completely changes the kinematics. The effect can be observed in Figure~\ref{fig:btau11}, where the distributions for the same variables as in Figure~\ref{fig:btau10} are shown. The same cuts are applied except  for $\Delta\phi_{\tau \nu}<0.4$, and the shown LQ signals correspond to the parameter choices $\beta_L^{33} = \beta_L^{23} = 1$, $\beta_R^{33} = 0$. As displayed in the upper panel of Figure~\ref{fig:btau11}, in the case of the~$m_T^{\tau}$~distribution the signal displays a peak at~$m_T^{\tau}=0$, corresponding to the $2\rightarrow2$ component, and a broad enhancement for $m_T^{\tau}> 300 \, {\rm GeV}$. The background is dominated by processes including the decay of a single $W$ boson, for which~$m_T^{\tau}$ is bounded at the parton level from above by the $W$-boson mass. A lower limit on $m_T^{\tau}$ of a few hundred~GeV is therefore useful to reduce the backgrounds. Notice that as a result of the different~$m_T^{\tau}$ selection and the missing $\Delta\phi_{\tau \nu}$ cut, the peak in the $m_{b \tau \nu}$ distribution corresponding to the $2 \to 2$ process is largely washed out  in signal samples that receive contributions from both  $b \tau \to U \to b \tau$ and  $g c \to U \nu_\tau \to b \tau \nu_\tau$ scattering. See the lower panel in Figure~\ref{fig:btau11} for an example distribution. 

The  selection $m_{b\tau}>160 \, {\rm GeV}$  imposed on the invariant mass of the $b$ and the $\tau_{\rm had}$ further suppresses the SM background including events results stemming from top-quark decays. An~additional  background rejection is achieved by requiring  $p_T(b)>100 \, {\rm GeV}$, $p_T(\tau_{\rm had})> 200 \, {\rm GeV}$, $\etmiss>400 \, {\rm GeV}$ and $\etmiss/p_T(\tau_{\rm had})<1.5$. The $p_T(b)$, the $\etmiss$ and the $\etmiss/p_T(\tau_{\rm had})$ distributions  after all cuts, besides the cut on the displayed variable, plus the requirement~$m_T^{\tau}> 400 \, {\rm GeV}$ are displayed in Figure~\ref{fig:btau11add}. In the case of the first two distributions one sees that the signal spectra fall off significantly slower than the corresponding background distributions. For the observable $\etmiss/p_T(\tau_{\rm had})$ the background is instead suppressed for low values of the ratio. The final selection is achieved by requiring and with a variable lower limit on  $m_T^{\tau}$ in the range between $400 \, {\rm GeV}$ and~$600 \, {\rm GeV}$. 

\subsection{Mono-top final state}
\label{sec:monotopanalysis}

\begin{figure}[t!]
\begin{center}
\includegraphics[width=0.55\textwidth]{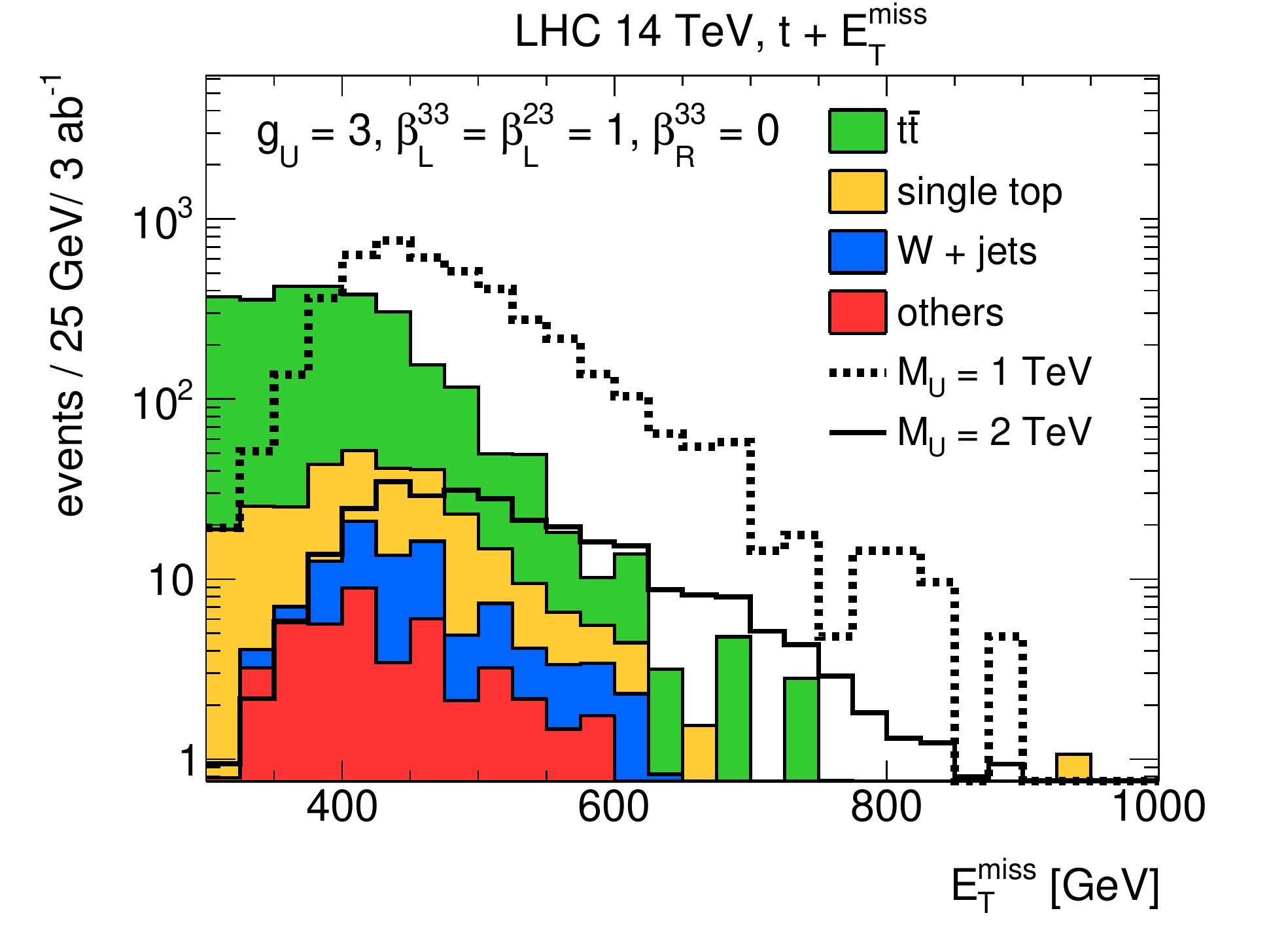}
\includegraphics[width=0.55\textwidth]{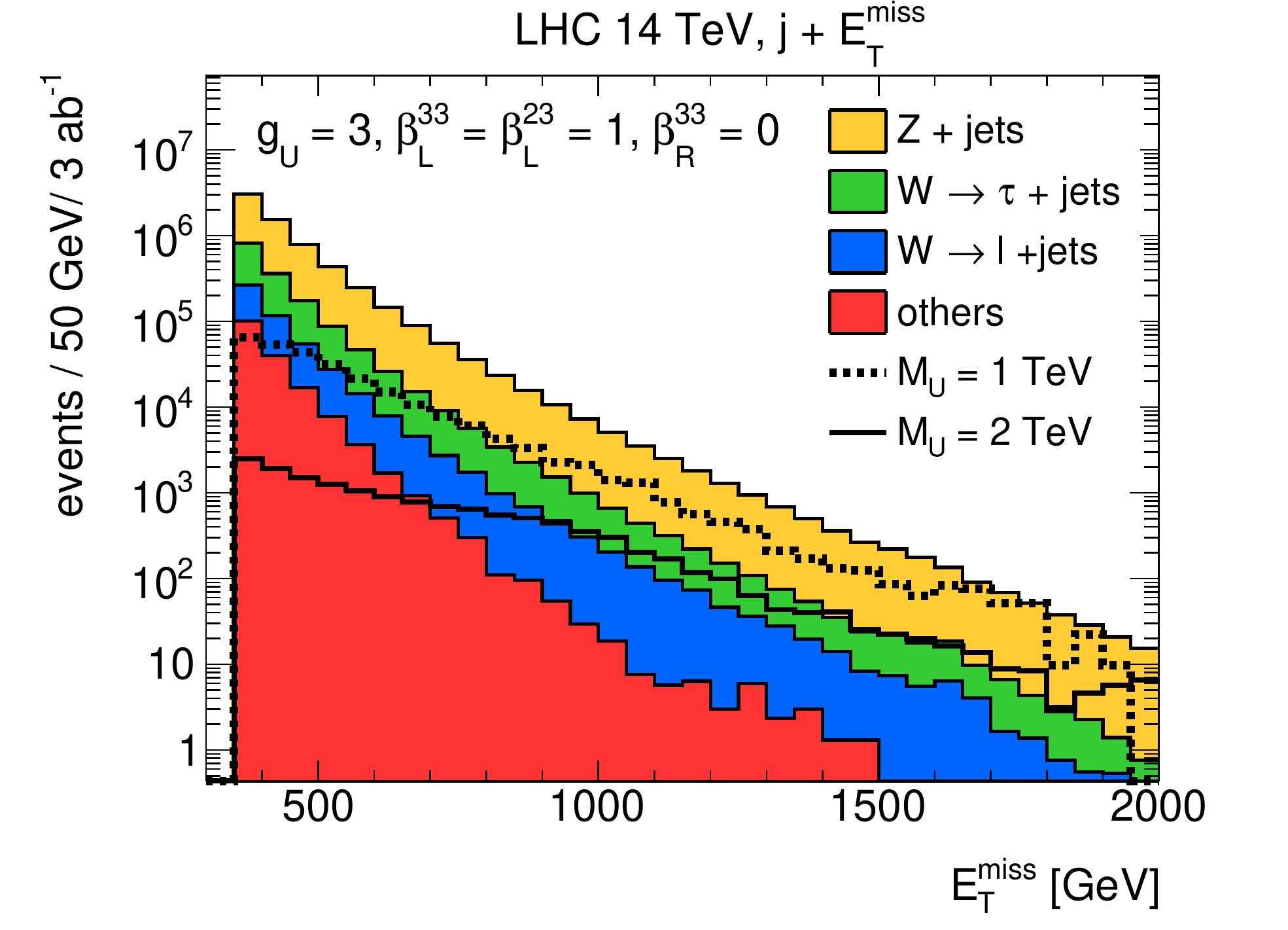}
\vspace{0mm}
\caption{Upper panel: $\etmiss$ distribution for the background and two $t + \etmiss$ signals. Lower panel:~$\etmiss$~distribution for the background  and two $j + \etmiss$  signals. The~background distributions (coloured histograms)  are stacked and the signal distributions in both panels correspond to  $g_U = 3$, $\beta_L^{33} = \beta_L^{23} = 1$, $\beta_R^{33} = 0$ and  $M_U = 1 \, {\rm TeV}$ (black dotted lines) or $M_U = 2 \, {\rm TeV}$~(black solid lines). The displayed predictions are obtained for LHC collisions at~$14 \, {\rm TeV}$.}
\label{fig:etmmonotopmonojet}
\end{center}
\end{figure}

The mono-top signature featuring the production of a top quark recoiling against an invisible particle has been studied at the LHC~\cite{Sirunyan:2018gka,Aaboud:2018zpr} in the framework of dark matter~(DM)  searches. Two types of top-quark decays can be used to analyse this signature, namely the semi-leptonic and the fully hadronic decay. Notice that for boosted top quarks such as those produced in the decay of a heavy~LQ targeting the fully hadronic final state  requires the reconstruction of hadronic decays of the $W$ boson through  jet substructure techniques. In order to avoid this complication, we concentrate in our work on the final state that arises  from semi-leptonic top decays which is easier to simulate with a simple parametrised technique. The final state of interest thus includes an isolated lepton~($e$~or~$\mu$) a $b$-tagged jet and $\etmiss$ associated to  two (three) neutrinos for the $2\rightarrow2$~($2 \to 3$) process. Although they lead to a somewhat different final-state kinematics the $2\rightarrow2$ and $2\rightarrow3$ mono-top processes cannot be cleanly separated as in the case of the $b + \tau$ final state. We therefore develop an analysis strategy valid for both cases and provide results with the two different topologies combined.

The basic selections in our mono-top analysis are one  lepton with $p_T(\ell)>30 \, {\rm GeV}$ and one $b$-jet with $p_T(b)>50 \, {\rm GeV}$ both within $|\eta|<2.5$. Additional leptons and $b$-jets are vetoed, and any additional light-flavour jet in the event is required to have $p_T(j)<50 \, {\rm GeV}$. The main handle to reject the SM background  is the transverse mass  $m_T^\ell$ built from the lepton momentum and $\etmiss$. The requirement $m_T^\ell>250 \, {\rm GeV}$ suppresses all  backgrounds with a single neutrino that result from a $W$-boson decay. The requirement $m_{b\ell}<140 \, {\rm GeV}$ with $m_{b\ell}$ the invariant mass  of the lepton and the $b$-jet ensures in addition that the two particles are compatible with the decay of a top quark.  In~order to guarantee that  the top quark is boosted, the sum of the transverse momenta of  the lepton and the $b$-jet is required to satisfy $p_T(\ell)+p_T(b)>400 \, {\rm GeV}$, and the angular distance between them to be $\Delta R_{\ell b} = \sqrt{\Delta\eta^2_{\ell b}+\Delta\phi^2_{\ell b}} <1$. Here $\Delta\phi_{\ell b}$ denotes the difference in azimuthal angle between the lepton and the $b$-jet. The final selection consists in setting a lower limit on the value of $\etmiss$ dependent on precise value of the LQ mass. The upper panel of Figure~\ref{fig:etmmonotopmonojet} shows the $\etmiss$ distribution   for the background and two mono-top signals after imposing all the discussed cuts.  

\subsection{Mono-jet final state}
\label{sec:monojetanalysis}

In the mono-jet case the relevant LQ signature consists of one high-transverse momentum $c$-jet and~$\etmiss$ associated to either one or two neutrinos in the case of the $2 \to 2$ or the $2 \to 3$ topology. Although $c$-jets can  be experimentally tagged at the~LHC~--- for actual ATLAS and CMS analyses relying on $c$-tagging see~\cite{Aaboud:2018fhh,Aaboud:2018zjf,Sirunyan:2019qia} ---  a high-transverse momentum mono-charm analysis has not been performed by the LHC experiments. We therefore do not attempt to exploit the fact that the jet in the LQ mono-jet signature results from a $c$ quark, but perform a flavour-blind $j +  \etmiss$ analysis to target  the $c \bar{\nu}_\tau \hspace{0.25mm} (\nu_\tau)$ final state.  Our analysis hence resembles  the canonical approach of searching for DM at the LHC, which has received much experimental~\cite{Aaboud:2016tnv,Aaboud:2017phn,Sirunyan:2017hci,ATLAS-CONF-2020-048} and theoretical~\cite{Lindert:2017olm} attention, resulting in  high-precision estimates of the dominant $\etmiss$ backgrounds associated to  the production of a~$Z$~or~$W$ boson accompanied by at least one high-transverse momentum jet. 

We use as a reference the ATLAS analysis described in~\cite{ATLAS-CONF-2020-048} but employ a higher $\etmiss$ cut of $\etmiss>350 \, {\rm GeV}$, which reflects the fact that we are aiming for the energetic decay products of a LQ with a mass in excess of $1 \, {\rm TeV}$. We require the presence of a high-transverse momentum jet with $p_T (j) >150 \, {\rm GeV}$ within $|\eta|<2.4$, and no more than four jets with $p_T (j)>30 \, {\rm GeV}$ within $|\eta|<2.8$. The selection $\Delta\phi_{\rm min}>0.4$, where $\Delta\phi_{\rm min}$ is the minimum angular difference in the azimuthal plane between a reconstructed jet and \etmiss, is used to fully suppress the multi-jet background. All events containing a reconstructed electron or muon, or the hadronic decay  of~a~$\tau$ are rejected. The sensitivity of the search is extracted through a multi-bin comparison of the shapes of the $\etmiss$ variable for the LQ signal  and the LQ signal plus the SM background. Our shape fit covers eight bins between $350 \, {\rm GeV}$ and $2 \, {\rm TeV}$ with an additional overflow bin for events with~$\etmiss>2 \, {\rm TeV}$. $\etmiss$ distributions   for the background and two mono-jet signals after applying the discussed selections are displayed in the lower panel of Figure~\ref{fig:etmmonotopmonojet}.

\section{Numerical results}
\label{sec:results}

On the basis of the selection criteria defined in Section~\ref{sec:strategies}, we will study the LHC sensitivity to the three LQ signatures discussed previously assuming integrated luminosities of $300 \, {\rm fb}^{-1}$ and $3 \, {\rm ab}^{-1}$. These luminosities correspond to the full statistics expected after LHC~Run~III and the HL-LHC phase, respectively.  For each signature and each point  in the LQ parameter space we evaluate the value of the cross section which  can be excluded at 95\%~confidence level (CL) normalised to the nominal  LO cross section for the relevant model realisation as calculated by {\tt MadGraph5\_aMC\@NLO}. The experimental sensitivity is evaluated using a test statistics based on a profiled likelihood ratio and we make use of the CLs method~\cite{Read:2002hq} as implemented in the {\tt RooStat} toolkit~\cite{Moneta:2010pm}.

\begin{figure}[t!]
\begin{center}
\includegraphics[width=0.475\textwidth]{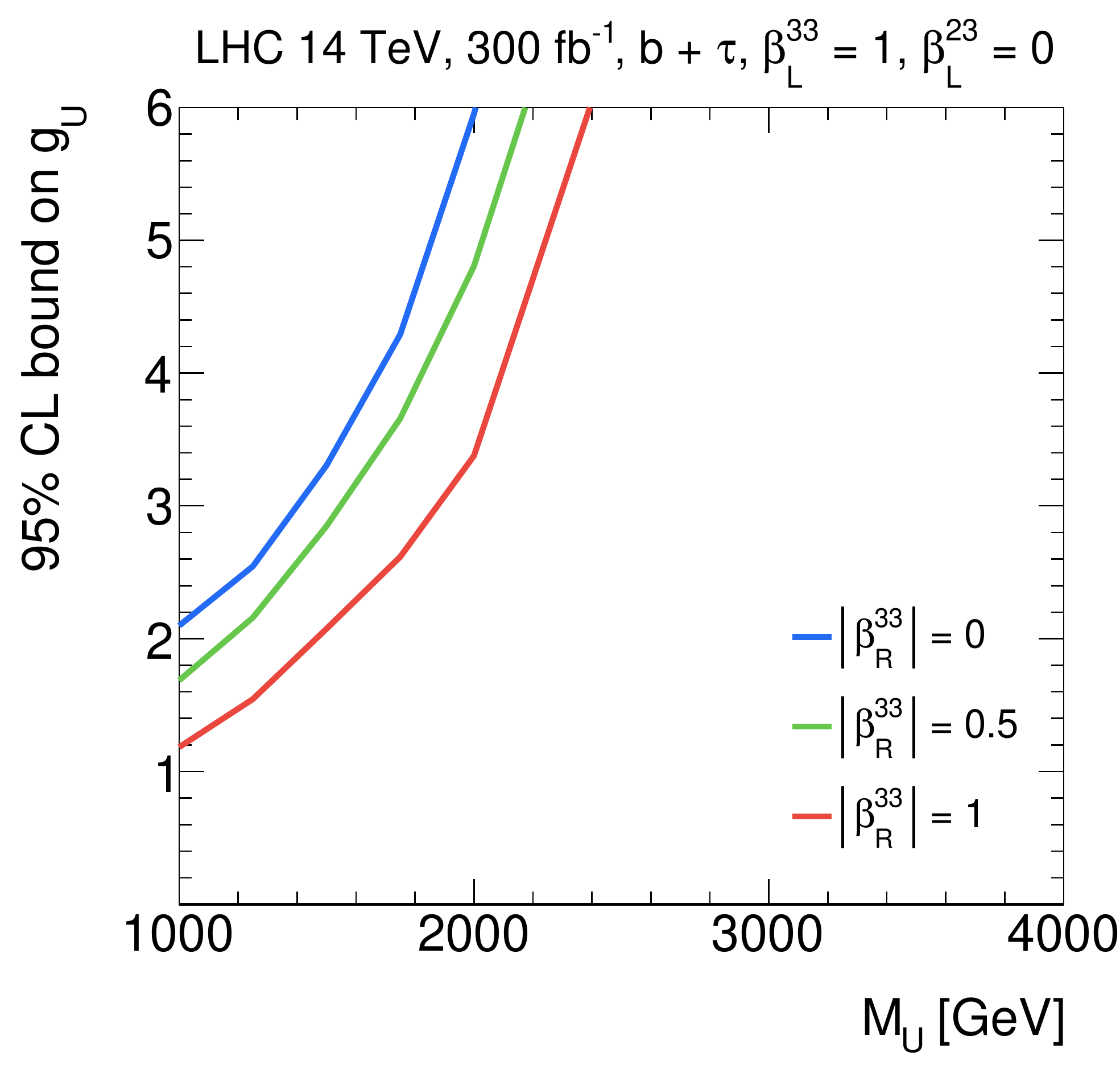} \quad 
\includegraphics[width=0.475\textwidth]{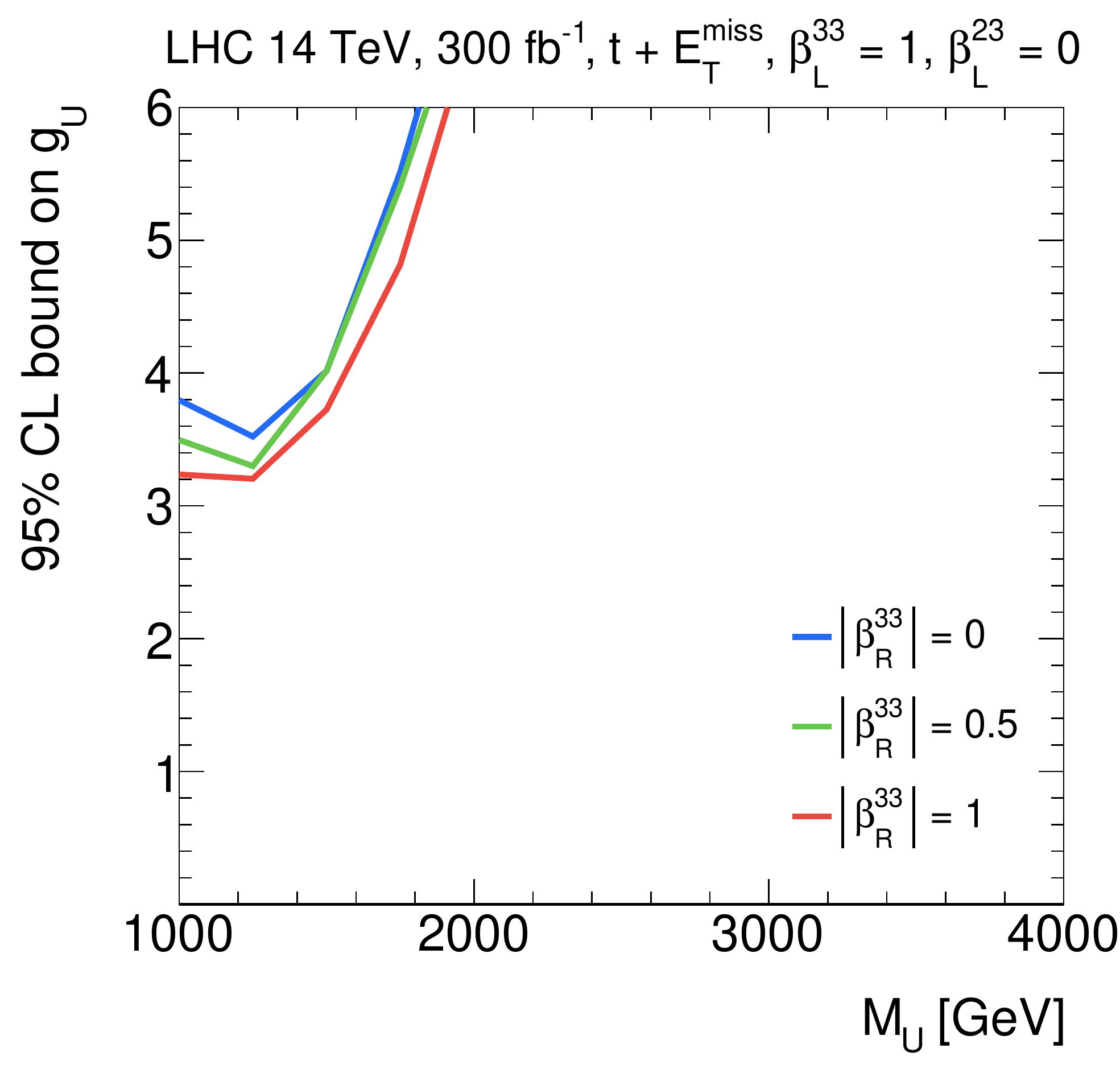}

\includegraphics[width=0.475\textwidth]{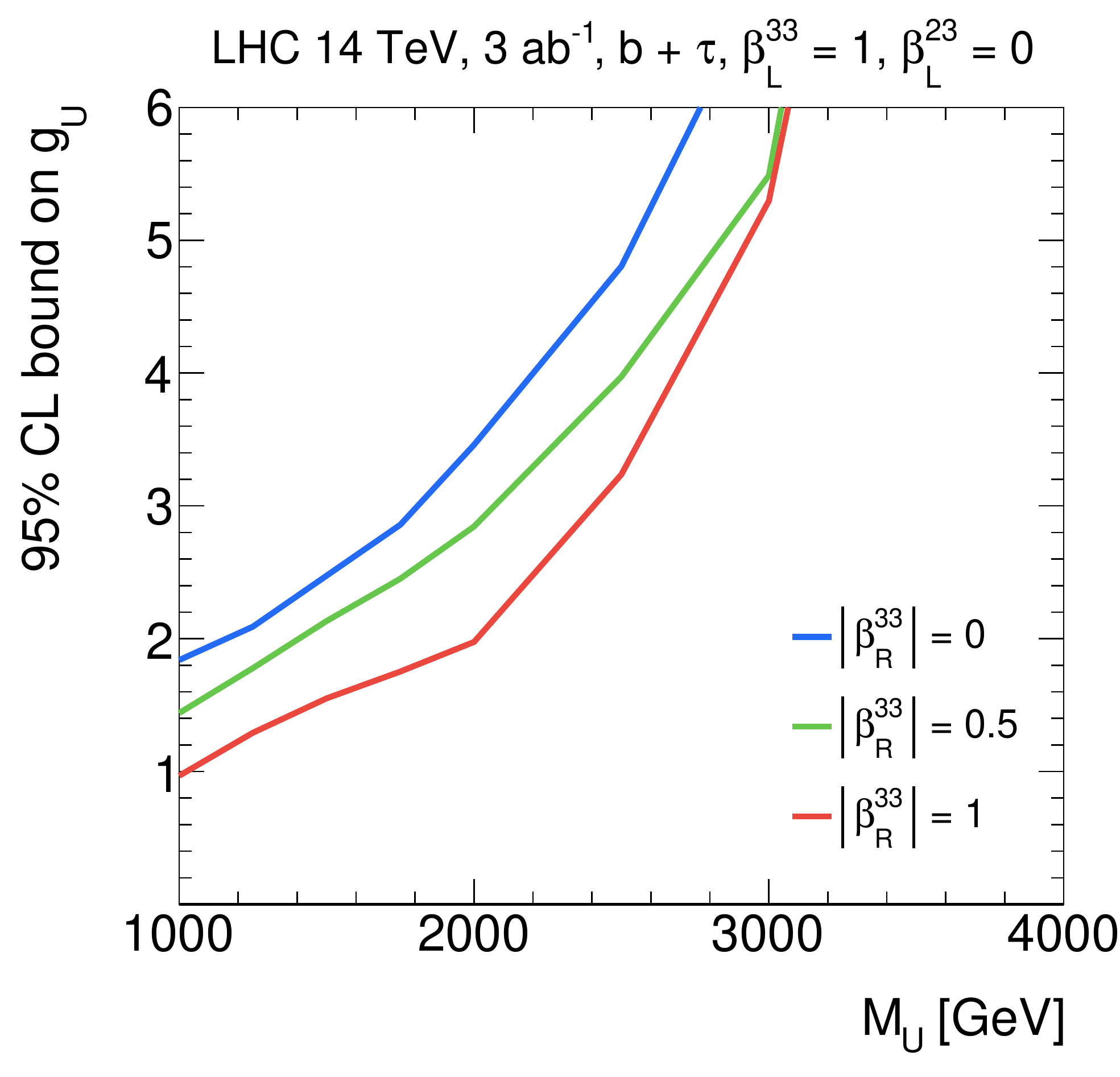} \quad 
\includegraphics[width=0.475\textwidth]{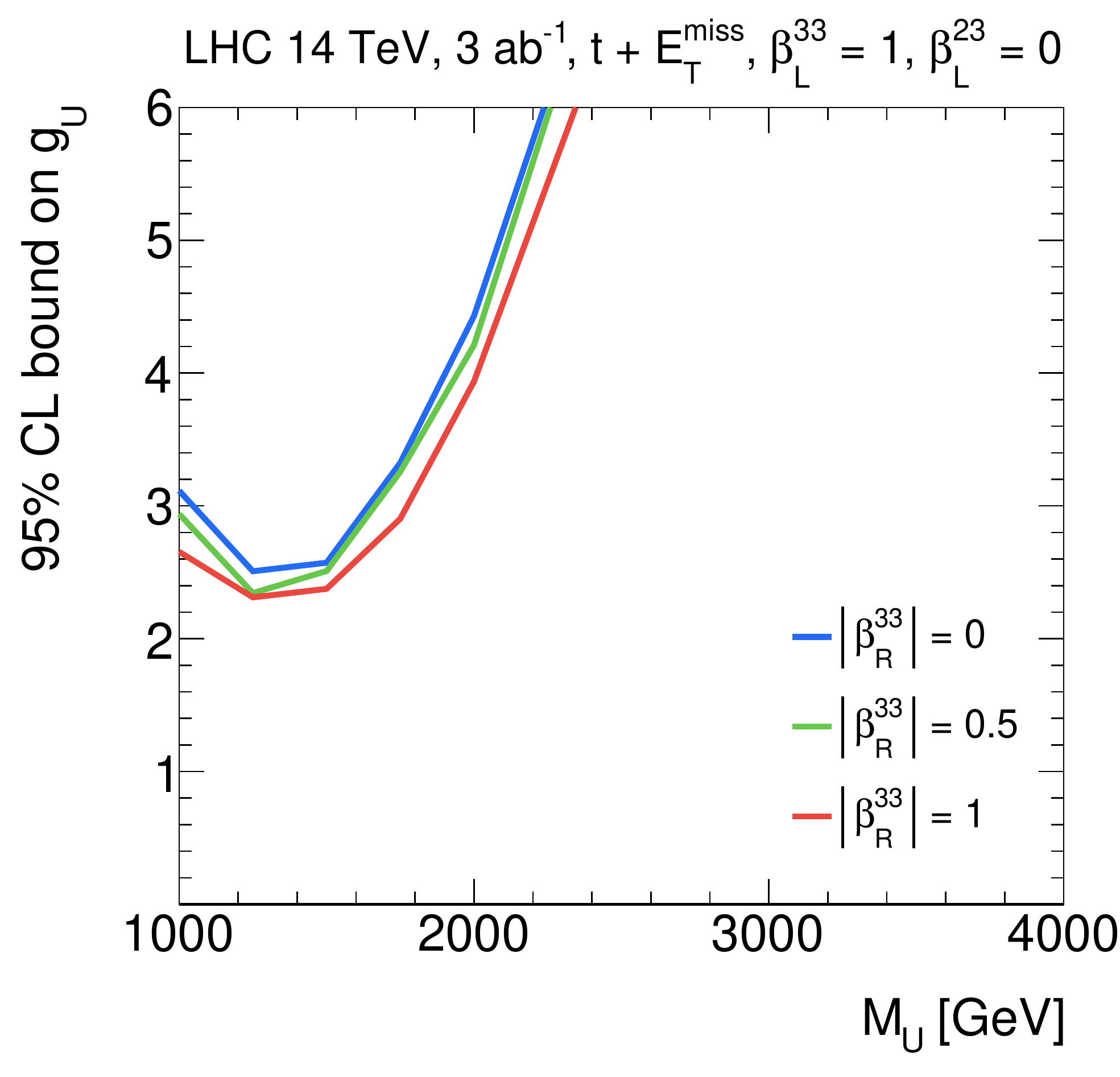}
\vspace{2mm}
\caption{95\%~CL exclusions in the $M_U\hspace{0.25mm}$--$\hspace{0.25mm}g_U$ plane following from the $b + \tau$ search~(left) and the $t + \etmiss$ search~(right).  The shown constraints employ the LQ parameters $\beta_L^{33} = 1$, $\beta_L^{23} = 0$ and $\left |\beta_R^{33} \right | = \{0, 0.5, 1\}$, and parameter regions above and to the left of the coloured lines are excluded. The upper (lower) panels assume $300 \, {\rm fb}^{-1}$ $\big ( 3 \, {\rm ab}^{-1} \big )$ of LHC collisions at $14 \, {\rm TeV}$.  }
\label{fig:MUgUplanes}
\end{center}
\end{figure}

\begin{figure}[t!]
\begin{center}
\includegraphics[width=0.475\textwidth]{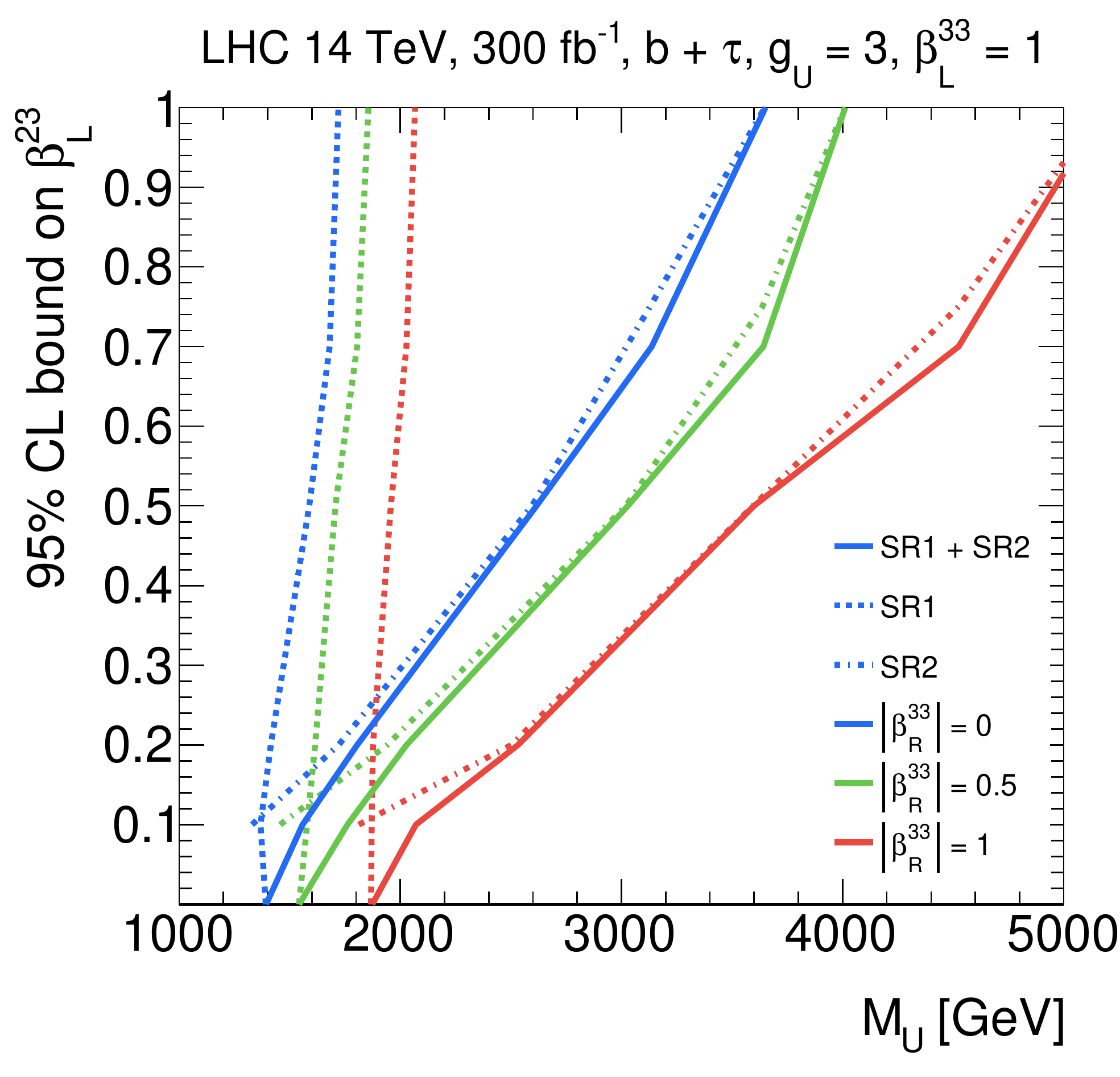} \quad 
\includegraphics[width=0.475\textwidth]{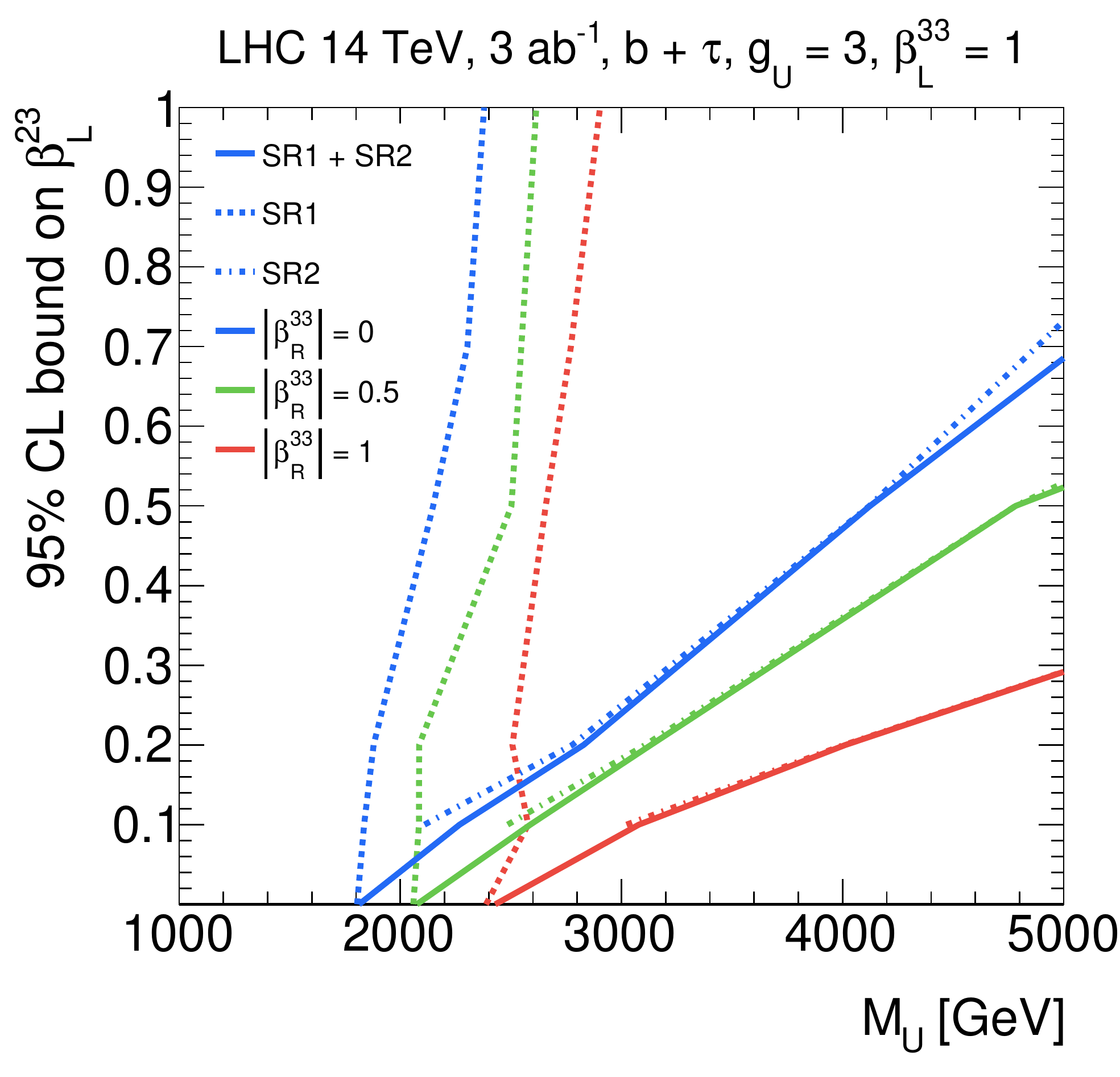}
\vspace{0mm}
\caption{95\%~CL exclusion following from the $b + \tau$ search in the $M_U\hspace{0.25mm}$--$\hspace{0.25mm}\beta_L^{23}$ plane assuming $300 \, {\rm fb}^{-1}$~(left) and $3 \, {\rm ab}^{-1}$~(right) of LHC $14 \, {\rm TeV}$ data. The parameter regions above and to the left of the coloured lines are excluded. The shown constraints are obtained for $g_U = 3$, $\beta_L^{33} = 1$ and $\left |\beta_R^{33} \right | = \{0, 0.5, 1\}$. The dotted, dash-dotted and solid lines illustrate the bounds arising from signal region one~(SR1), signal region two~(SR2) and their combination~(SR1 + SR2), respectively. }
\label{fig:btau23}
\end{center}
\end{figure}

In the case of the  $b + \tau$ and the mono-top signature, we assume systematic uncertainties of 15\% and 5\% on the SM backgrounds and on the signal, respectively.  As explained above, in our  mono-jet analysis we adopt a selection which closely resembles the signal region IM3 of the latest ATLAS $j + \etmiss$ search which is based on  $139 \, {\rm fb}^{-1}$ of LHC~Run~II data~\cite{ATLAS-CONF-2020-048}. The systematic uncertainty quoted by ATLAS in this signal region is 1.4\%, and we adopt this value as the systematic uncertainty on the total number of background events. Since we perform a fit to the shape of the $\etmiss$ distribution, we also need to take into account uncertainties related to the~$\etmiss$  shape. For each of the  $\etmiss$ bins considered in the analysis, ATLAS  provides an uncertainty which increases from approximately 1.4\% to 4\%  between $350 \, {\rm GeV}$ to $1.2 \, {\rm TeV}$. We apply these systematic uncertainties as bin-by-bin shape uncertainties in our fit.  For the bins between $1.5 \, {\rm TeV}$ and $2 \, {\rm TeV}$ we furthermore assume an uncertainty of $5\%$, while we take an uncertainty of 8\% for the total number of events with  $\etmiss > 2 \, {\rm TeV}$.  Notice that our uncertainty treatment corresponds to taking the uncertainties among different $\etmiss$ bins to be uncorrelated. In addition,  since the statistical uncertainties of the control regions that are used to constrain the background will get reduced with more luminosity, also the  systematic uncertainties are expected to decrease with larger data samples.  We therefore believe that  our mono-jet study provides conservative results when applied to the full LHC~Run~III and HL-LHC data sets. \\

In order to allow for an easy comparison with the comprehensive analysis~\cite{Baker:2019sli}  of LHC constraints on the singlet vector LQ model, we  adopt the choice $\beta_L^{33} = 1$, permit variations of $\beta_L^{23}$ and  consider~the following three benchmarks $\left |\beta_R^{33} \right | = \{0, 0.5, 1\}$ for the right-handed coupling $\beta_R^{33}$. We~begin our numerical study by considering LQ realisations~(\ref{eq:LU}) with $\beta_L^{23}=0$. In this case the dominant constraints on the parameter space~(\ref{eq:input}) arise from the proposed $b + \tau$ and  mono-top searches. The~95\%~CL exclusion limits in the $M_U\hspace{0.25mm}$--$\hspace{0.25mm}g_U$ plane that are obtained from the analyses described in~Sections~\ref{sec:btauanalysis} and \ref{sec:monotopanalysis} are shown in the panels of Figure~\ref{fig:MUgUplanes}.  To obtain the displayed results we have set~$\beta_L^{33} = 1$. From the panels it is evident that  the $b + \tau$ search generically leads to stronger constraints on $g_U$  than the mono-top search.  Furthermore, the limits from the $b + \tau$  search become notably  stronger  for non-zero~$\beta_R^{33}$, because both the $b \tau \to U$ production cross section and the~$U \to b \tau$ branching ratio are enhanced by~$\left | \beta_R^{33} \right|^2$.  In the mono-top case, the relevant~$U \to t \nu_\tau$ branching ratio is instead  proportional to~$1/\left | \beta_R^{33} \right|^{2}$ and as a result the derived constraints are to first approximation independent of~$\beta_R^{33}$. We furthermore observe that  for $M_U = 1 \, {\rm TeV}$ ($M_U = 3 \, {\rm TeV}$) the~95\%~CL limits on~$g_U$ that can be set at the HL-LHC is by a factor of around~$1.5$~($3.8$) better than the bound that LHC~Run~III is expected  to place. This~statement applies to both search strategies considered in the panels of Figure~\ref{fig:MUgUplanes}. Notice finally that for the largest values of $g_U$ shown in the latter figure, i.e.~$g_U \simeq 6$, higher-order radiative corrections are expected  to become numerically important, and as~a result  the derived limits are less reliable in this region.

So far we have  in this section only  studied LQ realisations with no second-third-generation left-handed mixing,~i.e.~$\beta_L^{23} = 0$. We now allow for $\beta_L^{23} \neq 0$ in which case one has to take into account the contributions from both the $2 \to 2$ and the $2 \to 3$ process~(see~Figure~\ref{fig:diagrams}).  In Figure~\ref{fig:btau23}~we show the~95\%~CL exclusion n the $M_U\hspace{0.25mm}$--$\hspace{0.25mm}\beta_L^{23}$ plane  that derive  from our $b + \tau$ search strategies. The corresponding  parameter choices are $g_U = 3$, $\beta_L^{33} = 1$ and $\left |\beta_R^{33} \right | = \{0, 0.5, 1\}$.  The~bounds arising from signal region one~(SR1), signal region two~(SR2) and their combination~(SR1~+~SR2) are indicated as dotted, dash-dotted and solid lines, respectively --- cf.~Section~\ref{sec:btauanalysis}~for the precise definition of the signal regions. SR1 targets the $2 \to 2$ process and one observes that the corresponding limits are largely independent from~$\beta_L^{23}$. This behaviour can be understood by noticing that for~$\beta_L^{23} \neq 0$ the LQ production  cross section receives contributions proportional to~$\left | \beta_L^{23} \right|^2$ and $\left | \beta_L^{13} \right|^2 \simeq \lambda^2 \left | \beta_L^{23} \right|^2 \simeq 0.05 \left | \beta_L^{23} \right|^2$  from the partonic quark-lepton channels~$s \tau \to U$ and~$d \tau \to U$, respectively. While the $d \tau \to U$ contribution is enhanced by the down-quark PDF it is also strongly~CKM  suppressed, and as a result the corresponding number of signal events in  SR1 amounts to no more than 10\% of the total rate. It furthermore turns out that~$b \tau \to U$  production provides the leading contribution to the number of events in SR1 for $\beta_L^{23} \lesssim 0.5$, while for  $\beta_L^{23} \gtrsim 0.5$ the $s \tau \to U$ channel dominates. For sufficiently large $\beta_L^{23}$ the LQ production cross section is thus enhanced by~$\left |\beta_L^{23} \right|^2$ but since the~$U \to b \tau$ branching ratio is  proportional to $1/\left |\beta_L^{23} \right|^{2}$ the limits on~$M_U$  become almost independent of $\beta_L^{23}$.

\begin{figure}[t!]
\begin{center}
\includegraphics[width=0.475\textwidth]{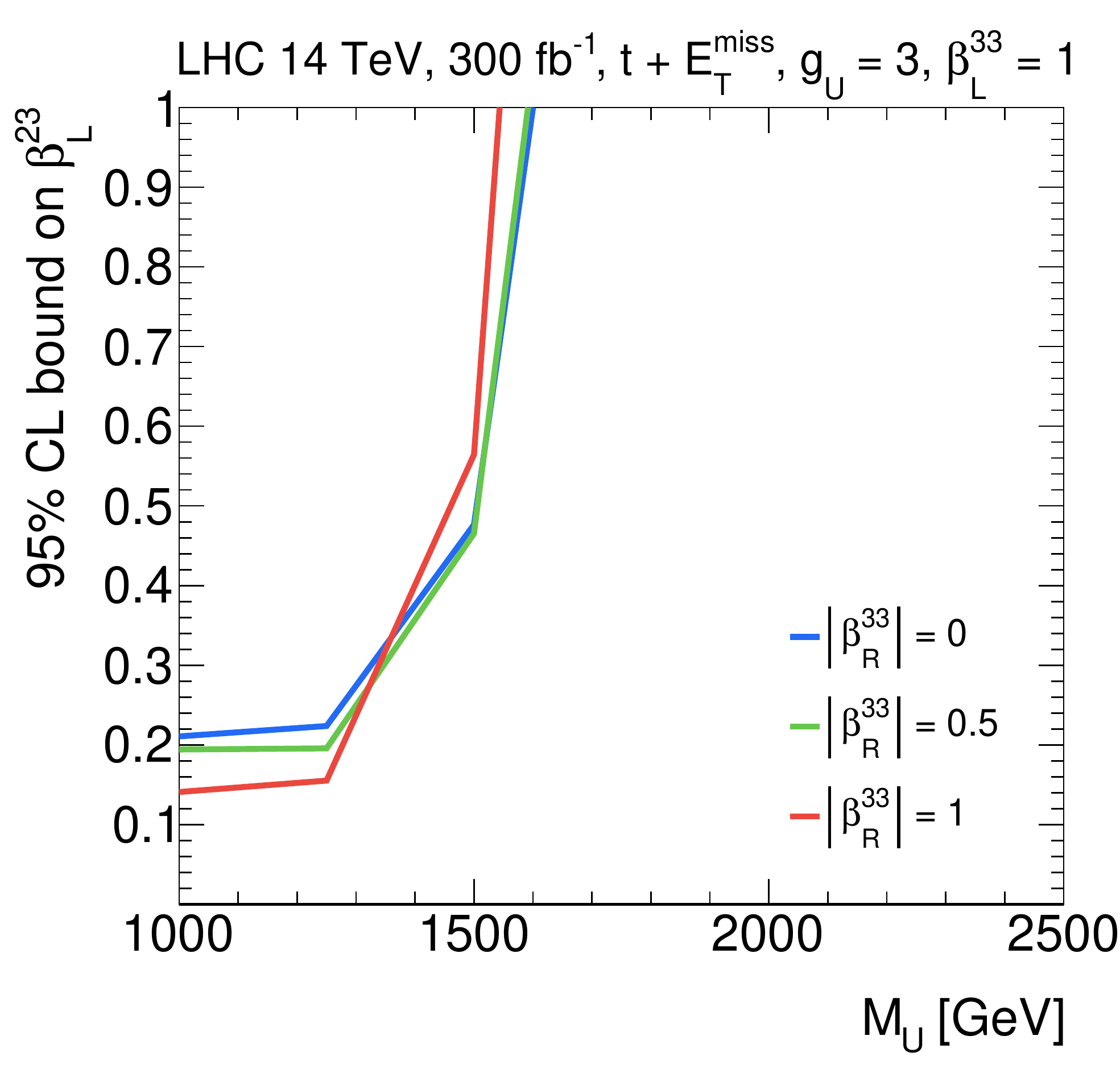} \quad 
\includegraphics[width=0.475\textwidth]{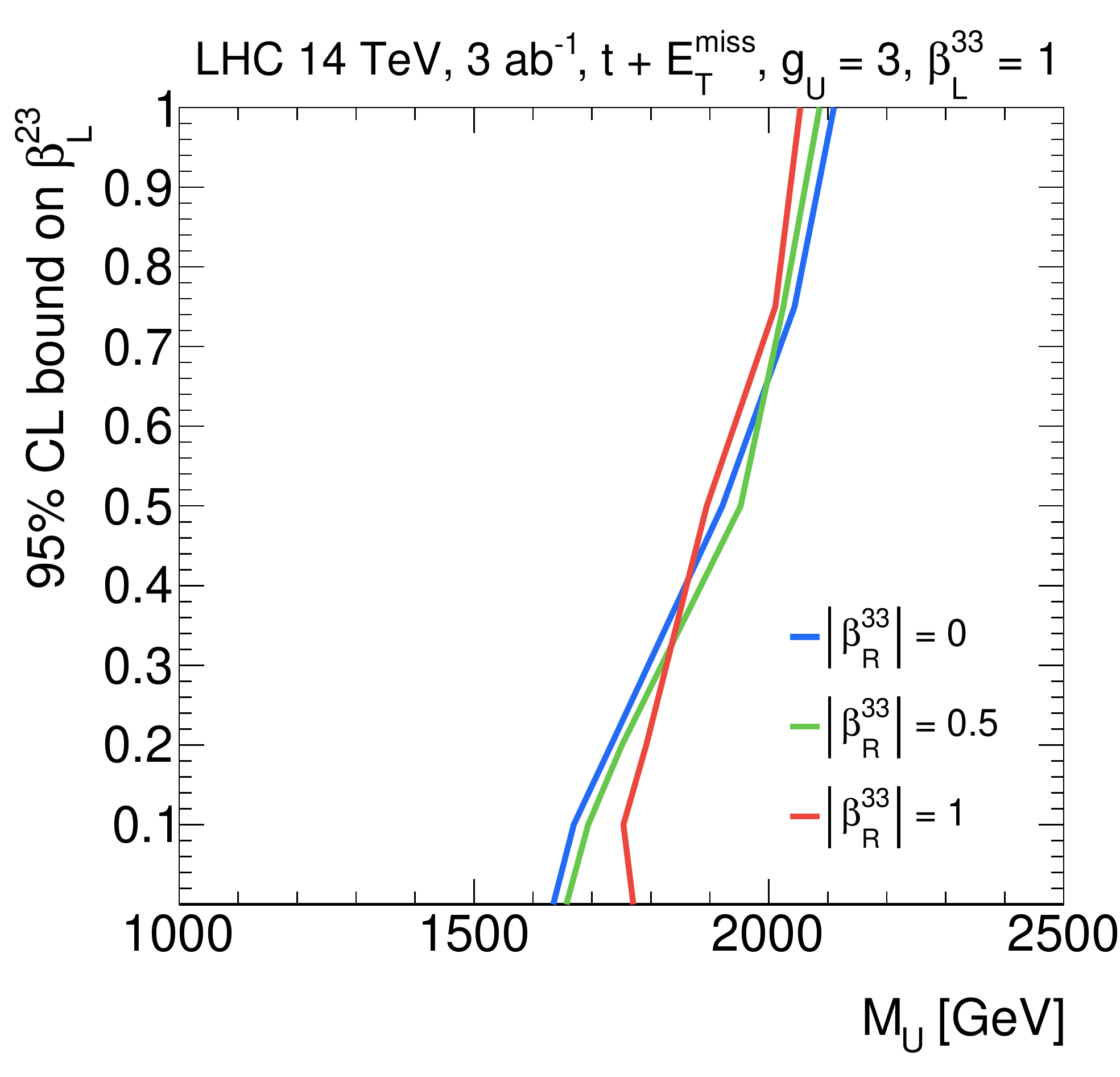}
\vspace{0mm}
\caption{As Figure~\ref{fig:btau23} but for the $t + \etmiss$ search. The shown exclusions correspond to the mono-top search including both the $2 \to 2$ and the $2 \to 3$ process.}
\label{fig:monotop23}
\end{center}
\end{figure}

\begin{figure}[t!]
\begin{center}
\includegraphics[width=0.475\textwidth]{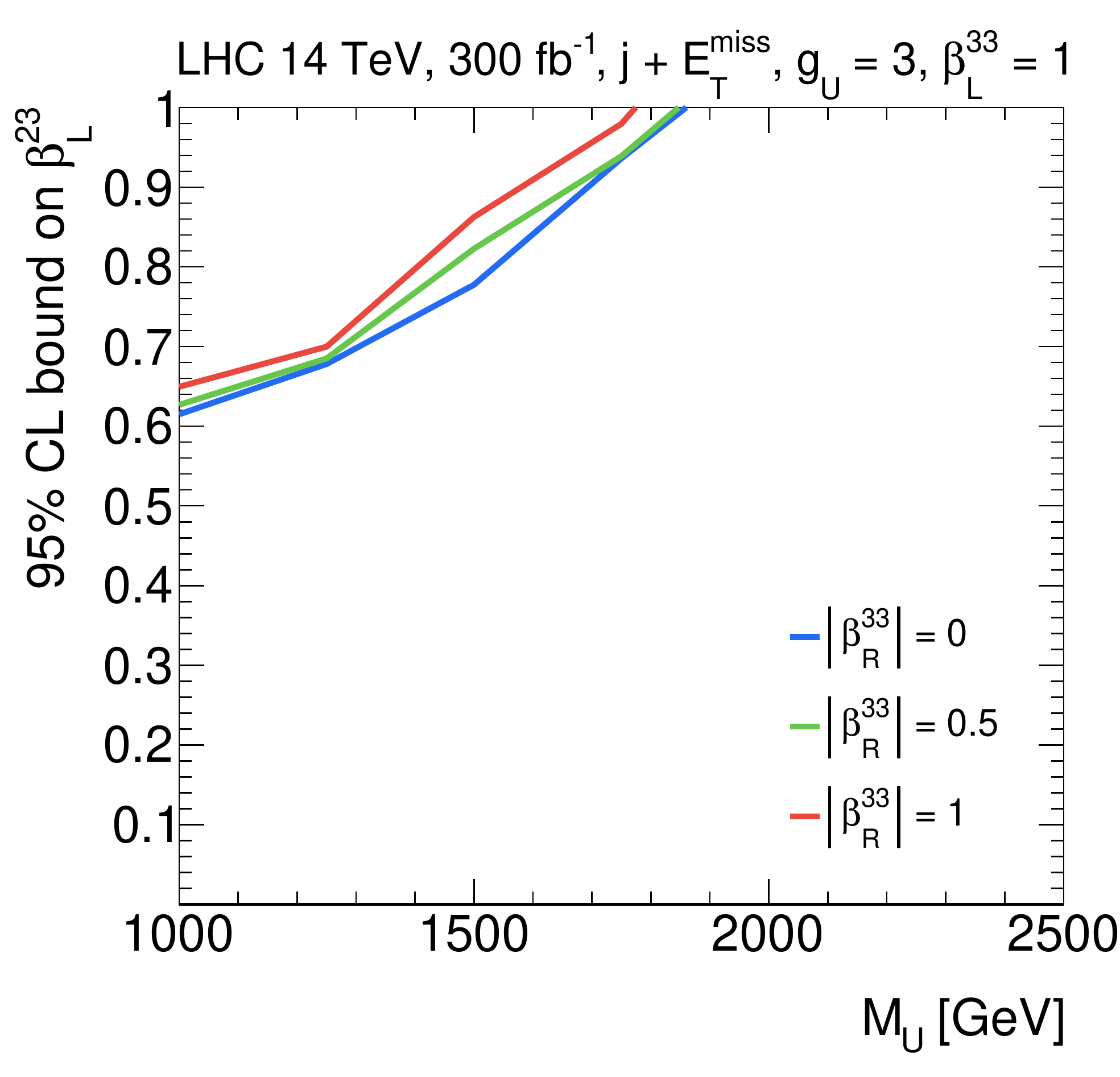} \quad 
\includegraphics[width=0.475\textwidth]{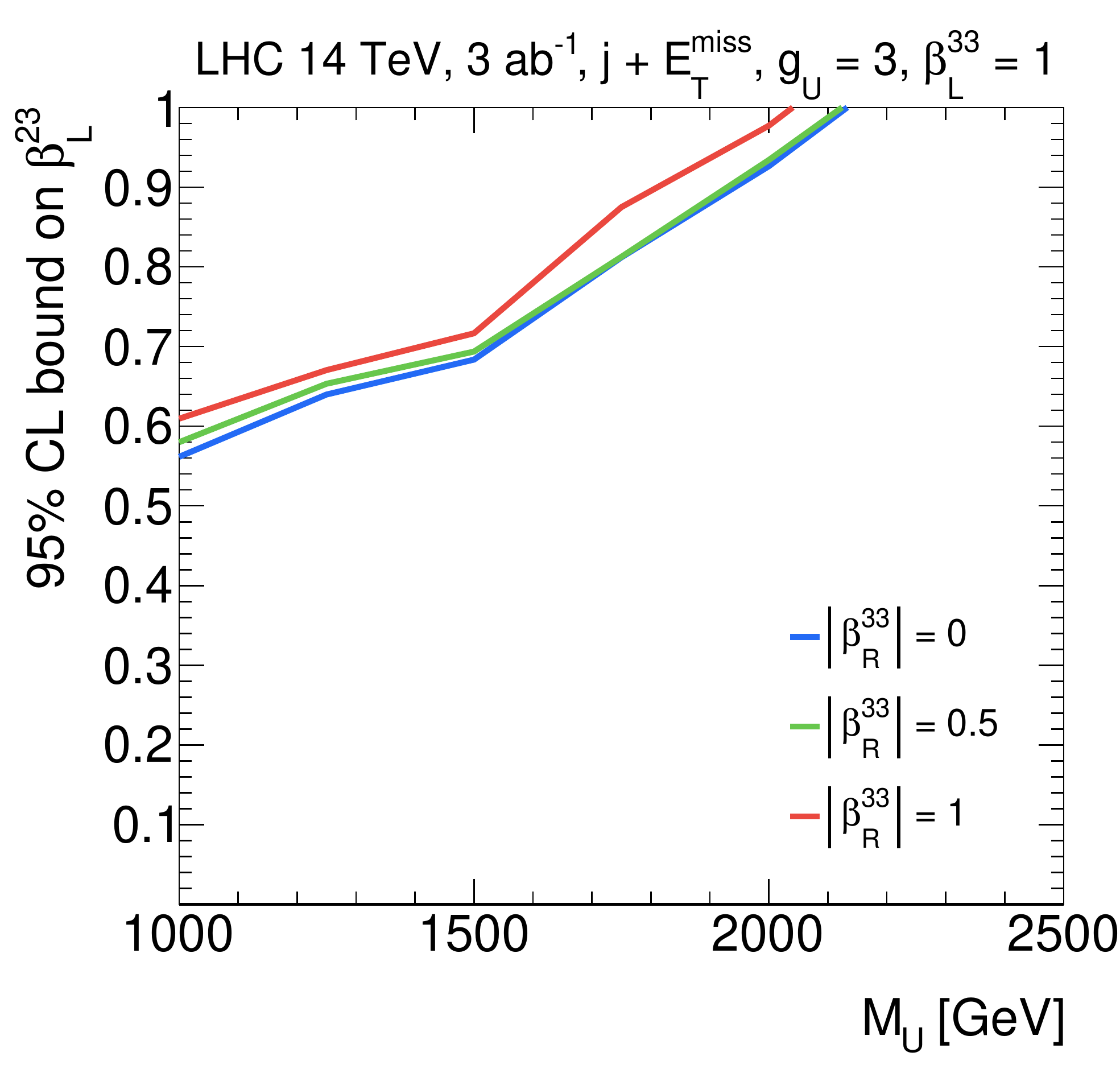}
\vspace{0mm}
\caption{As Figure~\ref{fig:monotop23} but for the $j + \etmiss$ search.  Both the $2 \to 2$ and the $2 \to 3$ process are included to obtain the displayed mono-jet exclusions.}
\label{fig:monojet23}
\end{center}
\end{figure}

Concerning the exclusions resulting from SR2,  we first remark that the  gluon-quark fusion channel $gc \to b \tau \nu_\tau$ ($gu \to b \tau \nu_\tau$)  contributes around 85\%~(15\%) of the number of events in this signal region. In fact, the largest contribution to the fiducial $gc \to b \tau \nu_\tau$ cross section arises from $t$-channel exchange of a $U$, which  turns out to scale approximately as $\beta_L^{23}$ after cuts. This finding explains why the constraints in the $M_U\hspace{0.25mm}$--$\hspace{0.25mm}\beta_L^{23}$ plane that follow from the $b + \tau$ search in SR2 are  roughly linear in $\beta_L^{23}$. Notice that the above arguments  hold for any value of the third-generation right-handed quark-lepton-LQ coupling, and in consequence the shape of the $b + \tau$ exclusions  in  Figure~\ref{fig:btau23} do not change under variations of $\beta_R^{33}$ but they  simply  become stronger with  increasing~$\left | \beta_R^{33} \right |$. It is also evident from  this figure that the~SR2 search strategy is significantly more powerful in constraining the  $M_U\hspace{0.25mm}$--$\hspace{0.25mm}\beta_L^{23}$ parameter space than the~SR1 search strategy. It~has to be stressed, however, that the $b + \tau$  final state resulting from the $2 \to 2$ process still plays a unique role, because in the case of a discovery, only this signature would potentially  allow to determine the LQ mass by observing a  peak in the~$m_{b \tau \nu_\tau}$ distribution at the nominal~LQ mass~$M_U$. 
 
In Figure~\ref{fig:monotop23} we display the 95\%~CL exclusions in the $M_U\hspace{0.25mm}$--$\hspace{0.25mm}\beta_L^{23}$ plane  that derive  from the  mono-top search strategy described in Section~\ref{sec:monotopanalysis}. The shown results correspond to the parameter choices $g_U = 3$, $\beta_L^{33} = 1$ and $\left |\beta_R^{33} \right | = \{0, 0.5, 1\}$. Like in the case of the panels on the right-hand side of~Figure~\ref{fig:MUgUplanes}, we see that the derived constraints are to first approximation independent of the  choice of $\beta_R^{33}$, because this dependence tends to cancel in the product of the production cross section and the branching ratio. One also observes that for the employed parameters it will not be possible to test  models with $\beta_L^{23} =0$ at LHC~Run~III  for any LQ~mass in the studied range. This is expected to change with higher integrated luminosities, and the HL-LHC mono-top searches  should be able to ultimately exclude models with $\beta_L^{23} =0$  up to~LQ~masses of~$M_U \simeq 1.7 \, {\rm TeV}$.  Concerning the~LHC~Run~III (HL-LHC) bounds, we mention that in the  LQ parameter space close to the exclusions around 60\% to 80\% (70\% to $100\%$) of the mono-top signal arises from the $2 \to 2$~process. We furthermore  observe that for $\beta_L^{23} \lesssim 0.5$ the dominant mono-top production  mechanism is  $b \tau \to U \to t \nu_\tau$, while for  $\beta_L^{23} \gtrsim 0.5$ the $s \tau \to U \to t \nu_\tau$ channel provides the leading contribution. The channel $d \tau \to U \to t \nu_\tau$ is always subleading, amounting to no more than $15\%$ of the total~$2 \to 2$ rate.  In the case of the $2 \to 3$ process, we  find instead that around 85\% of the events arise from $g c \to U \nu_\tau  \to t \nu_\tau \bar \nu_\tau$, while $15\%$ are due to $g u \to U \nu_\tau  \to t \nu_\tau \bar \nu_\tau$.

The 95\%~CL exclusions in the $M_U\hspace{0.25mm}$--$\hspace{0.25mm}\beta_L^{23}$ plane  that arise from   the mono-jet analysis  described in~Section~\ref{sec:monojetanalysis} can finally be seen in Figure~\ref{fig:monojet23}.  To allow for a straightforward comparison with the $b+\tau$ and mono-top limits  discussed before,  we have again chosen $g_U = 3$, $\beta_L^{33} = 1$ and $\left |\beta_R^{33} \right | = \{0, 0.5, 1\}$. Since the dependence on the third-generation right-handed coupling  tends to cancel in the product of the production cross section and the branching ratio the obtained mono-jet limits are again largely insensitive to the choice of $\beta_R^{33}$. One also observes that compared to the $b + \tau$ and the mono-top search, the reach of the mono-jet search is significantly weaker. In the considered LQ scenario   it should be possible to exclude values $\beta_L^{23} \gtrsim 0.6$ for $M_U = 1 \, {\rm TeV}$, while for $\beta_L^{23} =  1$ the LHC seems to have a mass reach of $M_U \lesssim 2 \, {\rm TeV}$. Also notice that the shown HL-LHC exclusions are only slightly stronger than the limits obtained at LHC~Run~III. This is a result of the conservative treatment of systematic uncertainties in our mono-jet study. 

\section{Conclusions and outlook}
\label{sec:conclusions} 

The main goal of this article was to give additional well-motivated examples of lepton-initiated processes  that can provide valuable probes of beyond  the SM physics  at hadron colliders once the lepton PDFs  are precisely known~\cite{Buonocore:2020nai}. In contrast to the recent detailed study~\cite{Buonocore:2020erb} which considered minimal scalar LQs coupling to first- and second-generation leptons and quarks, we have analysed the LHC reach of resonant production modes involving third-generation singlet vector~LQs. Such particles are enjoying a renaissance because they can simultaneously explain the  deviations from  $\tau$-$\mu$ (and $\tau$-$e$)  universality seen in $b \to c \ell \nu$ transitions and the deviations from~$\mu$-$e$  universality observed in $b \to s \ell^+ \ell^-$ transitions. 

We have argued that the observed $B$ anomalies if interpreted in the context of a  third-generation singlet vector LQ model strongly motivate searches for resonant LQ production in the $b + \tau$, the mono-top and the mono-jet final state. While the latter two final states have been targeted at the LHC (see~for instance~\cite{Sirunyan:2018gka,Aaboud:2018zpr,Sirunyan:2017jix,ATLAS-CONF-2020-048}), searches for  final states with a single bottom quark and a single tau  lepton have up till now not been performed by either  ATLAS nor CMS. In the case of the~$b + \tau$ final state, we have designed two different analyses strategies that aim at separating the $2 \to 2$ from the  $2 \to 3$ process~(cf.~Figure~\ref{fig:diagrams}) by exploiting the fact that the resulting final states  have very different kinematic features. Although they lead to a somewhat different final-state kinematics, in the mono-top and the mono-jet case the $2\rightarrow2$ and $2\rightarrow3$ processes cannot be cleanly separated. We  have therefore developed analysis strategies valid for both cases and provide mono-top and mono-jet results with the two different topologies combined.

\begin{figure}[t!]
\begin{center}
\includegraphics[width=0.475\textwidth]{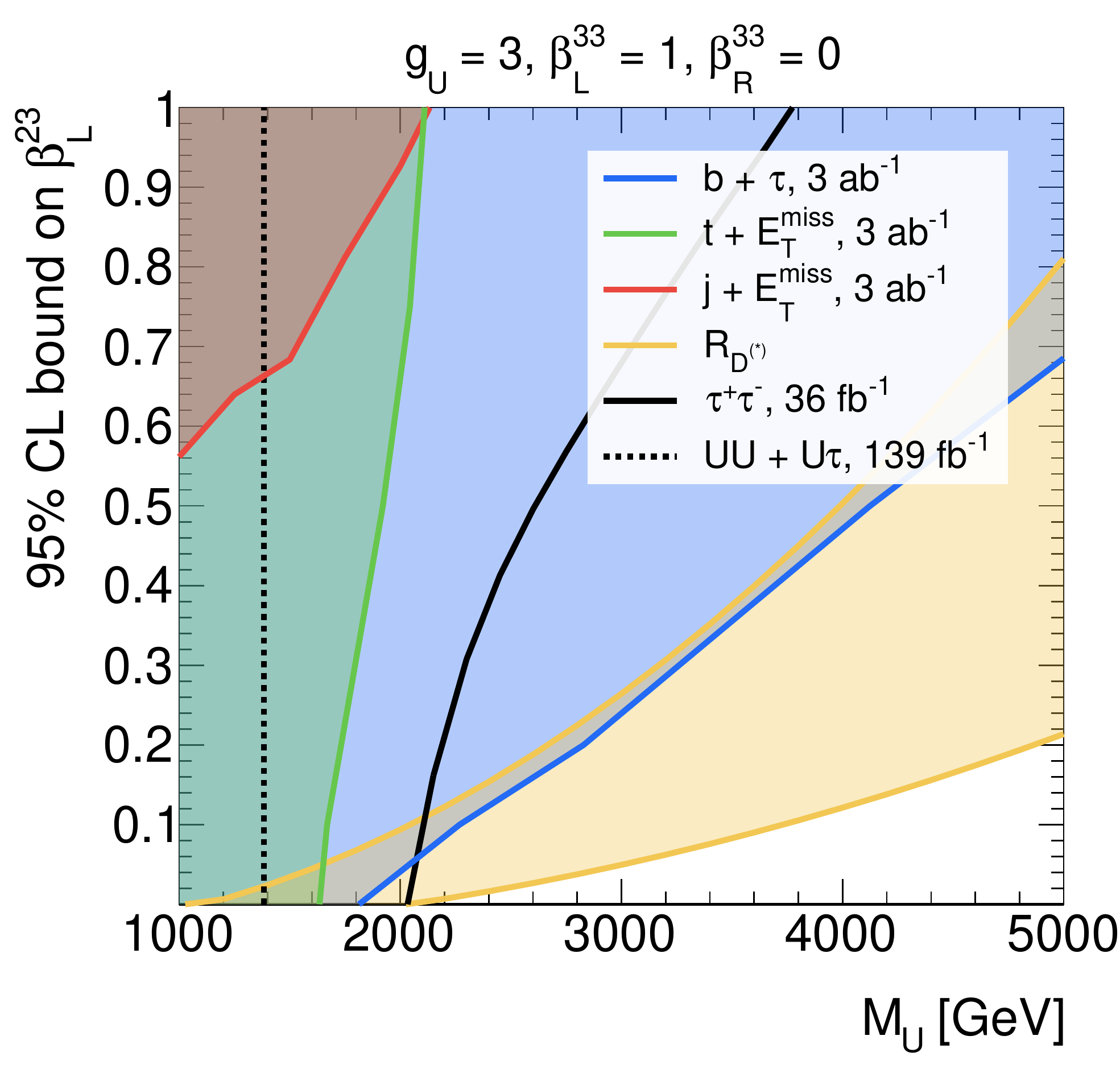} \quad 
\includegraphics[width=0.475\textwidth]{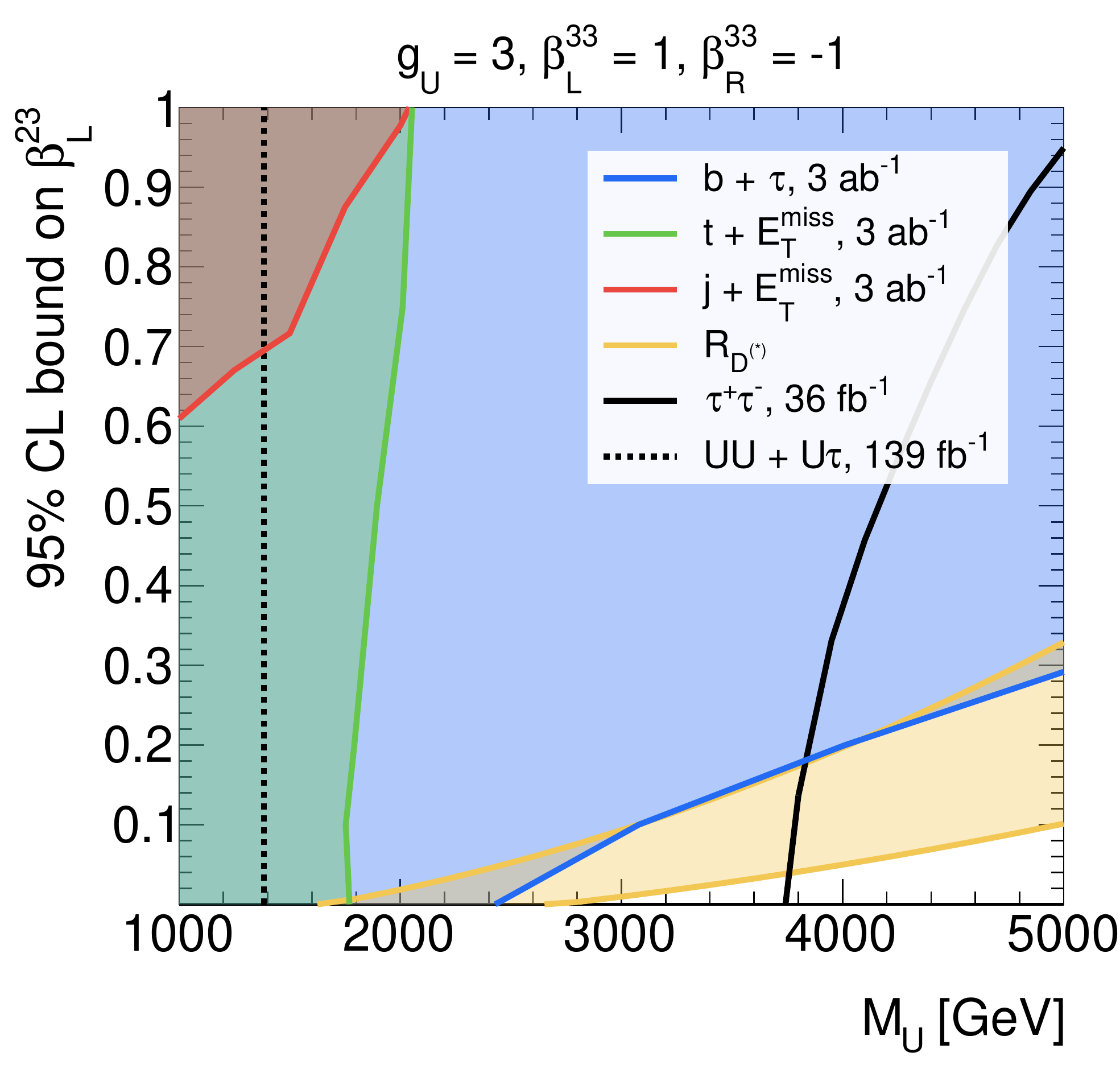}
\vspace{0mm}
\caption{Current and future  constraints on the  third-generation  singlet vector LQ  in the $M_U\hspace{0.25mm}$--$\hspace{0.25mm}\beta_L^{23}$ plane. The left~(right) panel corresponds to the parameter choices $g_U = 3$, $\beta_L^{33} = 1$ and $\beta_R^{33} = 0$~$\big($$\beta_R^{33} = -1$$\big)$. The~shown blue, green and red regions indicated the 95\%~CL exclusions  that derive from the proposed $b + \tau$,  $t + \etmiss$ and $j + \etmiss$ search strategies assuming $3 \, {\rm ab}^{-1}$ of LHC $14 \, {\rm TeV}$ data. For comparison also the regions favoured by the $b \to c\ell \nu$ data~$(R_{D^{(\ast)}})$, the bounds that have been derived in~\cite{Baker:2019sli}  from  di-tau~($\tau^+ \tau^-$) production and the limits following from the recent CMS search for pair and single third-generation LQ signatures~$(UU + U \tau)$~\cite{cms_2012.04178} are indicated as yellow contours,  solid black curves and dotted black lines, respectively. The displayed  $\tau^+ \tau^-$ ($UU + U \tau$) constraints correspond to $36 \, {\rm fb}^{-1}$ $\big($$139 \, {\rm fb}^{-1}$$\big)$ of LHC $13 \, {\rm TeV}$ data. See text for further details. }
\label{fig:comparison}
\end{center}
\end{figure}

\begin{figure}[t!]
\begin{center}
\includegraphics[width=0.475\textwidth]{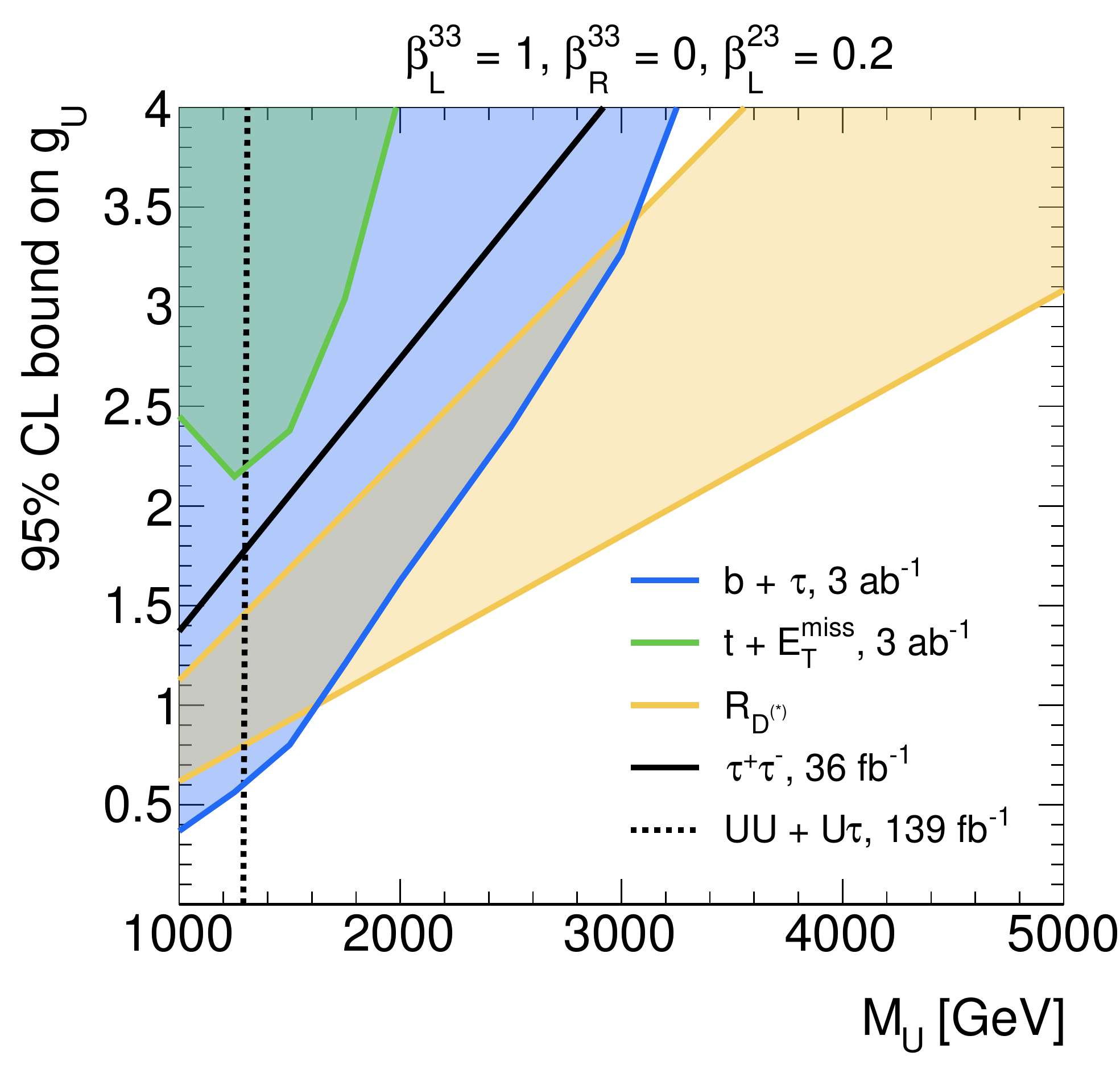} \quad 
\includegraphics[width=0.475\textwidth]{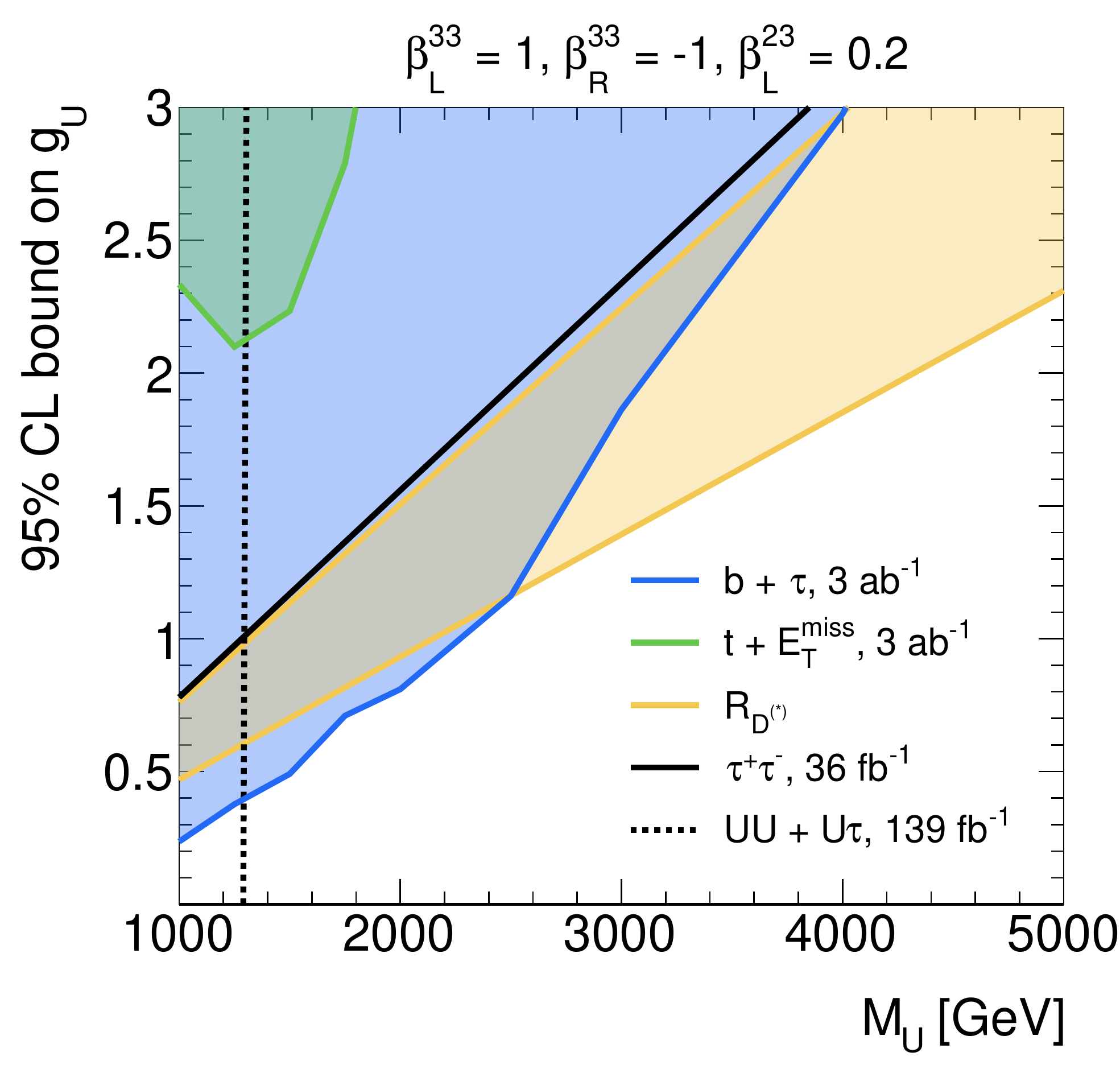}
\vspace{0mm}
\caption{Current and future  constraints on the  third-generation  singlet vector LQ  in the $M_U\hspace{0.25mm}$--$\hspace{0.25mm}g_U$ plane. The~left~(right) panel corresponds to the benchmarks  $\beta_L^{33} = 1$, $\beta_R^{33} = 0$~$\big($$\beta_R^{33} = -1$$\big)$ and $\beta_L^{23} = 0.2$. The~colour and style coding of the shown constraints are the same as
those displayed in~Figure~\ref{fig:comparison}. Since for $\beta_L^{23} = 0.2$, the proposed $j + \etmiss$ search strategy does  lead to bounds $g_U > 4$  even at the HL-LHC with $3 \, {\rm ab}^{-1}$ of integrated luminosity, mono-jet constraints do not appear in the two panels.}
\label{fig:summary}
\end{center}
\end{figure}

We have then analysed the coverage of the parameter space of the third-generation singlet  vector LQ model at LHC~Run~III and the HL-LHC. In a first step, we have studied model realisations with 
no second-third-generation left-handed mixing,~i.e.~$\beta_L^{23} = 0$, considering the three benchmarks $\left |\beta_R^{33} \right | = \{0, 0.5, 1\}$ for the third-generation right-handed coupling. In this case the dominant 95\%~CL constraints in the $M_U\hspace{0.25mm}$--$\hspace{0.25mm}g_U$ plane  arise from the $b + \tau$ and the mono-top search. In a second step, we have then allowed for $\beta_L^{23} \neq 0$ fixing the remaining parameters to $g_U = 3$ and $\beta_L^{33} = 1$. The two panels in Figure~\ref{fig:comparison} illustrate  the HL-LHC coverage  of the proposed $b + \tau$,  mono-top and mono-jet search strategies in the $M_U\hspace{0.25mm}$--$\hspace{0.25mm}\beta_L^{23}$ plane for $\beta_R^{33} = 0$~(left) and $\beta_R^{33} = -1$~(right).
For comparison the~95\%~CL regions preferred by a fit to the LFU violating ratios $R_{D^{(\ast)}}$~(see~e.g.~\cite{Aebischer:2019mlg,Baker:2019sli,Cornella:2019hct}~for details), the $pp \to \tau^+ \tau^-$ limits obtained in~\cite{Baker:2019sli} and the bounds from the recent CMS~search~\cite{cms_2012.04178} for $pp \to UU$ and $pp \to U \tau$ production are also shown as yellow contours, solid black curves and dotted black lines, respectively. Concerning the latter bound of $M_U \simeq 1.4 \, {\rm TeV}$, we stress that it has been obtained under the assumption that the   vector LQ decays with  equal branching ratios of 50\% to both $b\tau^+$ and $t \bar \nu_\tau$ final states. Since~this assumption  does not hold  in the entire parameter space considered in the panels of Figure~\ref{fig:comparison} the shown $UU + U \tau$ exclusions have only an indicative character. From the plots in the latter figure it is evident that future LHC searches for a $b + \tau$ signature will allow to exclude  a relevant portion of the parameter space preferred by the $B$-physics anomalies in several models with third-generation singlet  vector LQs.  In order to further illustrate the latter statement, we show in Figure~\ref{fig:summary} the HL-LHC coverage of our $b + \tau$ search strategy in the $M_U\hspace{0.25mm}$--$\hspace{0.25mm}g_U$~plane  for $\beta_L^{33} = 1$ and $\beta_L^{23} = 0.2$. This is a representative benchmark to simultaneously address both the $R_{D^{(\ast)}}$ and the $R_{K^{(\ast)}}$ anomalies. The~left and right panel corresponds to   $\beta_R^{33} = 0$ and $\beta_R^{33} = -1$, respectively. For comparison the 95\% CL constraints following from the proposed mono-top search strategy~(green~contours), $R_{D^{(\ast)}}$ (yellow~contours), $pp \to \tau^+ \tau^-$~(solid black lines) and $pp\to  UU + U \tau$ production~(dotted black lines) are shown as well. One observes that for $\beta_R^{33} = 0$ ($\beta_R^{33} = -1$), HL-LHC searches for $b \tau$ final states should allow to test third-generation singlet vector~LQ explanations of the $B$-physics anomalies for LQ masses up to  around $3 \, {\rm TeV}$ $\big($$4 \, {\rm TeV}$$\big)$.  We~thus  encourage  the ATLAS and the CMS collaboration to perform dedicated resonance searches in final states featuring a single bottom quark   and a single tau lepton, a signature that has so far  not been explored by them. 

The results presented in this article also call for further theoretical developments concerning the accurate predictions of  LHC processes that are induced by quark-lepton annihilation. While~the lepton PDFs are now accurately known, official PS codes  can at present not correctly handle incoming leptons. In~addition, fully differential NLO~QCD and QED hadron-collider predictions are not available for processes like $b \tau \to U  \to b \tau$ --- a first step into this direction has  been made very recently in~\cite{Greljo:2020tgv} by calculating the NLO~QCD and QED corrections to the inclusive   quark-lepton induced cross sections of singly-produced scalar  LQs. Once fully differential NLO~QCD and QED corrections are available, they should be  consistently   combined with the PS to obtain~NLO+PS predictions that provide an accurate modelling of the exclusive final states targeted at the LHC by~ATLAS~and~CMS. 

\acknowledgments UH thanks Luca~Buonocore, Paolo~Nason, Francesco~Tramontano and Giulia~Zanderighi  for an enjoyable collaboration on a related topic.

\end{document}